\documentclass[twocolumn,tighten]{aastex61}
\usepackage{amssymb, amsfonts, amsmath,framed}

\usepackage{bm}
\expandafter\ifx\csname package@font\endcsname\relax\else
 \expandafter\expandafter
 \expandafter\usepackage
 \expandafter\expandafter
 \expandafter{\csname package@font\endcsname}
\fi
\expandafter\ifx\csname package@font\endcsname\relax\else
 \expandafter\expandafter
 \expandafter\usepackage
 \expandafter\expandafter
 \expandafter{\csname package@font\endcsname}
\fi
\def\bq{\begin{equation}}
\def\eq{\end{equation}}
\def\bqy{\begin{eqnarray}}
\def\eqy{\end{eqnarray}}

\def\calo{\mathcal{O}}

\begin{document}
\title{Plasmoid Instability in Forming Current Sheets}

\correspondingauthor{Luca Comisso}
\email{lcomisso@princeton.edu}

\author{L. Comisso}
\affiliation{Department of Astrophysical Sciences, Princeton University, Princeton, New Jersey 08544, USA}
\affiliation{Princeton Plasma Physics Laboratory, Princeton University, Princeton, New Jersey 08543, USA}

\author{M. Lingam}
\affiliation{Harvard-Smithsonian Center for Astrophysics, Cambridge, Massachusetts 02138, USA}
\affiliation{John A. Paulson School of Engineering and Applied Sciences, Harvard University, Cambridge, Massachusetts 02138, USA}

\author{Y.-M. Huang}
\affiliation{Department of Astrophysical Sciences, Princeton University, Princeton, New Jersey 08544, USA}
\affiliation{Princeton Plasma Physics Laboratory, Princeton University, Princeton, New Jersey 08543, USA}

\author{A. Bhattacharjee}
\affiliation{Department of Astrophysical Sciences, Princeton University, Princeton, New Jersey 08544, USA}
\affiliation{Princeton Plasma Physics Laboratory, Princeton University, Princeton, New Jersey 08543, USA}

\begin{abstract}
The plasmoid instability has revolutionized our understanding of magnetic reconnection in astrophysical environments. By preventing the formation of highly elongated reconnection layers, it is crucial in enabling the rapid energy conversion rates that are characteristic of many astrophysical phenomena. Most of the previous studies have focused on Sweet-Parker current sheets, which, however, are unattainable in typical astrophysical systems. Here, we derive a general set of scaling laws for the plasmoid instability in resistive and visco-resistive current sheets that evolve over time. Our method relies on a principle of least time that enables us to determine the properties of the reconnecting current sheet (aspect ratio and elapsed time) and the plasmoid instability (growth rate, wavenumber, inner layer width) at the end of the linear phase. After this phase the reconnecting current sheet is disrupted and fast reconnection can occur. The scaling laws of the plasmoid instability are \emph{not} simple power laws, and depend on the Lundquist number ($S$), the magnetic Prandtl number ($P_m$), the noise of the system ($\psi_0$), the characteristic rate of current sheet evolution ($1/\tau$), as well as the thinning process. We also demonstrate that previous scalings are inapplicable to the vast majority of the astrophysical systems. We explore the implications of the new scaling relations in astrophysical systems such as the solar corona and the interstellar medium. In both these systems, we show that our scaling laws yield values for the growth rate, wavenumber, and aspect ratio that are much smaller than the Sweet-Parker based scalings.
\end{abstract}

\section{Introduction} \label{SecIntro}
It is now generally acknowledged that magnetic reconnection powers some of the most important and spectacular astrophysical phenomena in the Universe such as coronal mass ejections, stellar flares, non-thermal signatures of pulsar wind nebulae, and gamma-ray flares in blazar jets \citep{TS97,Kuls05,ZwYa09,BG10,JD11,KZ15,Kag15}. Although the importance of magnetic reconnection has been recognized since the 1950s, it has recently witnessed an upsurge in popularity due to the realization of the importance of the plasmoid instability in facilitating fast reconnection (and energy release).

\begin{figure}[b]
\begin{center}
\vspace{0.60cm}
\includegraphics[width=8.6cm]{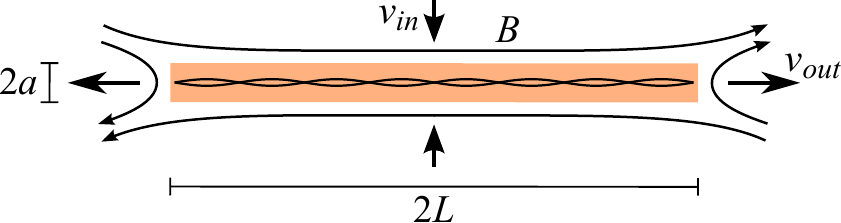}
\end{center}
\caption{Sketch of linear plasmoids forming in a reconnecting current sheet. The shaded orange region indicates the out-of-plane current sheet whose total width and length are $2a$ and $2L$, respectively. Plasmoids are represented by the thin magnetic islands in the current sheet.}
\label{fig01}
\end{figure}

The plasmoid instability can be understood in terms of a tearing instability occurring in a reconnecting current sheet (see Fig. \ref{fig01}). Numerical simulations providing clear indications that thin reconnecting current sheets may be unstable to the formation of plasmoids date back at least to the 1980s  \citep{Bisk82,Steinolf1984,ML85,Bisk86,LeeFu1986}, but it is only in the last decade that their role in speeding up the reconnection process has been widely appreciated. Indeed, very narrow reconnection layers would form in the absence of the plasmoid instability, which, in turn, would have the effect of throttling the reconnection rate. However, reconnecting current sheets exceeding a certain aspect ratio cannot form because they become unstable to the formation of plasmoids, which break the reconnection layer into shorter elements, consequently leading to a significant increase in the reconnection rate \citep{Daugh2006,Daugh2009}. Hence, the predictions of the classical Sweet-Parker reconnection model \citep{Sweet1958,Parker1957} break down for sufficiently large systems such as those typically encountered in astrophysical environments - in these cases, it was shown that the reconnection rate becomes nearly independent of the magnetic diffusivity   \citep{BHYR,HB10,ULS10,LSSU12,HB13,Taka2013,CGW15,Ebrahimi2015,ComGra16}.

The ability of plasmoid-mediated reconnection to enable fast energy release has been exploited in explaining multiple phenomena in a wide range of astrophysical 
settings with considerable success. They include solar flares \citep{ShiTa01,Barta2011a,Barta2011b,Li2015,ShiTa16,Jan17}, coronal mass ejections \citep{Mill10,Karpen12,Ni12,Mei2012,LMS15}, chromospheric jets \citep{Shi07,NKLW15,NZML17}, blazar emissions \citep{Giannios2013,SPG15,PGS16,Belob2017}, gamma-ray bursts \citep{Gian10,McU12,KZ15} and non-thermal signatures of pulsar wind nebulae \citep{SirSpi14,Guo2015,Werner2016,SGP16,Guo16}. 
Plasmoid formation can also produce self-generated turbulent reconnection \citep{Daughton2011,Oishi2015,HuBha16,Wang2016,Kowal2016}, implying that large scale current sheets are likely to become turbulent during the advanced stages of the reconnection process \citep{DeDe16}. Given the importance of nonlinear plasmoids in the reconnection process, several studies have also been devoted to the understanding of their statistical properties \citep{FDS10,ULS10,HB12,GBH13,Shen2013,JDD14,SGP16,Lynch2016,Almonte2016,LCB17}, which may be crucial to understand the occurrence of large abrupt events in solar, stellar and other massive objects flares \citep{ShiM11}.

Although the impact of plasmoids in reconnection has been thoroughly documented, there are many fundamental issues that still remain unresolved. Several of them have to do with the linear phase of the plasmoid instability, for which a comprehensive dynamical picture is still missing. The linear phase of the instability is of  fundamental importance because it allows us to understand in which conditions, and at what time, fast reconnection (which occurs when the plasmoids enter the nonlinear phase) is triggered. It is the goal of this paper to advance our theoretical understanding of the plasmoid instability in astrophysically relevant plasmas, but a historical background is first necessary to place our work in context with previous theoretical studies. 

It is rather intriguing that the first derivation of the  growth rate $\gamma$ and the wavenumber $k$ for a special case of the plasmoid instability was presented in an exercise of a textbook. Indeed, this textbook by \citet{TS97} showed that if one assumes that the reconnecting current sheet has an inverse-aspect-ratio corresponding to the Sweet-Parker one, $a/L \sim S^{-1/2}$, then the growth rate and the wavenumber of the plasmoid instability scale as $\gamma \tau_A \sim S^{1/4}$ and $kL \sim S^{3/8}$, respectively, where $S := L v_A/ \eta$ is the Lundquist number and $\tau_A := L/v_A$ is the Alfv{\'e}nic timescale, being $\eta$ the magnetic diffusivity and $v_A$ the Alfv{\'e}n speed calculated upstream of the current sheet. Therefore, the growth rate and the wavenumber (proportional to the number of plasmoids) increase monotonically with the Lundquist number of the system.

Surprisingly, these scaling relations remained overlooked until they were independently rederived by \citet{LSC07} a decade later. They were also generalized further through the inclusion of three-dimensional \citep{BBH12}, Hall \citep{Baalrud2011} and plasma viscosity \citep{LSU13,ComGra16} effects. However, it soon became apparent that the results obtained by assuming a Sweet-Parker current sheet were problematic because of the growth rate being proportional to $S$ raised to a positive exponent, which implies that the growth rate approaches infinity in the limit $S \rightarrow \infty$. For sufficiently high $S$-values, such as those typically encountered in astrophysical environments, the predicted growth rate would be so fast that the plasmoids would reach the nonlinear phase and disrupt the current sheet before the Sweet-Parker aspect ratio can be attained. This, of course, implies that a Sweet-Parker current sheet may not be realizable in the first place.

In order to circumvent this problem, \citet{PV14} conjectured that reconnecting current sheets break up when the condition $\gamma {\tau_A} \sim 1$ is met \footnote{The same condition was used to propose visco-resistive \citep{TRVP} and collisionless \citep{DSP16,Pucci2017} scalings.} (more precisely $\gamma {\tau_A} \simeq 0.623$). This led them to an alternative set of scalings, including the result that the final inverse-aspect-ratio in the resistive regime depends only on the Lundquist number $S$ and corresponds to $a/L \simeq S^{-1/3}$. A similar criterion was adopted in \citet{UL16}, who posited that the linear stage of the plasmoid instability essentially ends when $\gamma \tau = 1$, with $\tau$ being the characteristic time scale of current sheet evolution. However, each of these assumptions are open to question. If one `terminates' the linear dynamics at this stage, the growth rate is only comparable to the timescale of the current sheet thinning, implying that one cannot use the static dispersion relations of the tearing instability to carry out the calculations. Second, at this stage, the effects of the reconnection layer outflow are also non-negligible, since the modes are subject to stretching.

These difficulties are mostly rendered void when one observes that typically $\gamma \tau \gg 1$ at the end of the linear phase. On top of that, it is also important to note that the most interesting dynamics occurs when $\gamma > 1/\tau$. This has been shown in a recent Letter published by us \citep{CLHB16}, where it was demonstrated that the plasmoid instability exhibits a quiescence period followed by a rapid growth. Furthermore, the scaling relations of the plasmoid instability were shown to be \emph{no longer simple power laws}, as they included non-negligible logarithmic contributions and also depended upon the \emph{noise of the system}, the \emph{characteristic rate of current sheet evolution}, and even the nature of the \emph{thinning process}. This has direct implications for the onset of fast magnetic reconnection, because the correct identification of the scaling laws of the plasmoid instability is necessary for understanding when and how plasmoids becomes nonlinear and disrupt the reconnecting current sheet. 

In this work, we extend the analysis presented in the aforementioned Letter by formulating a detailed treatment of both the inviscid and viscous regimes of the plasmoid instability. A proper treatment of the latter is very important since viscosity (or equivalently, the magnetic Prandtl number, defined below) plays a major role in several astrophysical systems like accretion discs around neutron stars and black holes \citep{BH08}, warm interstellar medium \citep{BraSub2005}, protogalactic plasmas \citep{Kulsrud1997} and intergalactic medium \citep{Subra2006}. Our work accords four major advantages over prior studies: (i) the scaling laws for the plasmoid instability in general time-evolving current sheets are derived both in the resistive and visco-resistive regimes, (ii) a clear demarcation of the limited domain in which the previous scalings are applicable, (iii) the presentation of accurate results in astrophysically relevant regimes with very high $S$-values, and (iv) the exploration of the astrophysical implications of the plasmoid instability in the stellar and interstellar medium contexts.

The outline of the paper is as follows. In Sec. \ref{SecLeastTime}, the least time principle, which is used to compute the properties of the dominant mode at the end of the linear phase, is introduced. This is followed by a derivation of the resistive and visco-resistive scaling laws for the plasmoid instability in Secs. \ref{SecResistive} and \ref{SecViscoRes}, respectively. We discuss the astrophysical relevance of the derived scaling relations by choosing two systems (the solar corona and the interstellar medium) in Sec. \ref{Discuss}. Finally, we summarize our results in Sec. \ref{SecConc}.

\section{Least time principle for plasmoids} \label{SecLeastTime}
\begin{figure}
\begin{center}
\includegraphics[width=8.6cm]{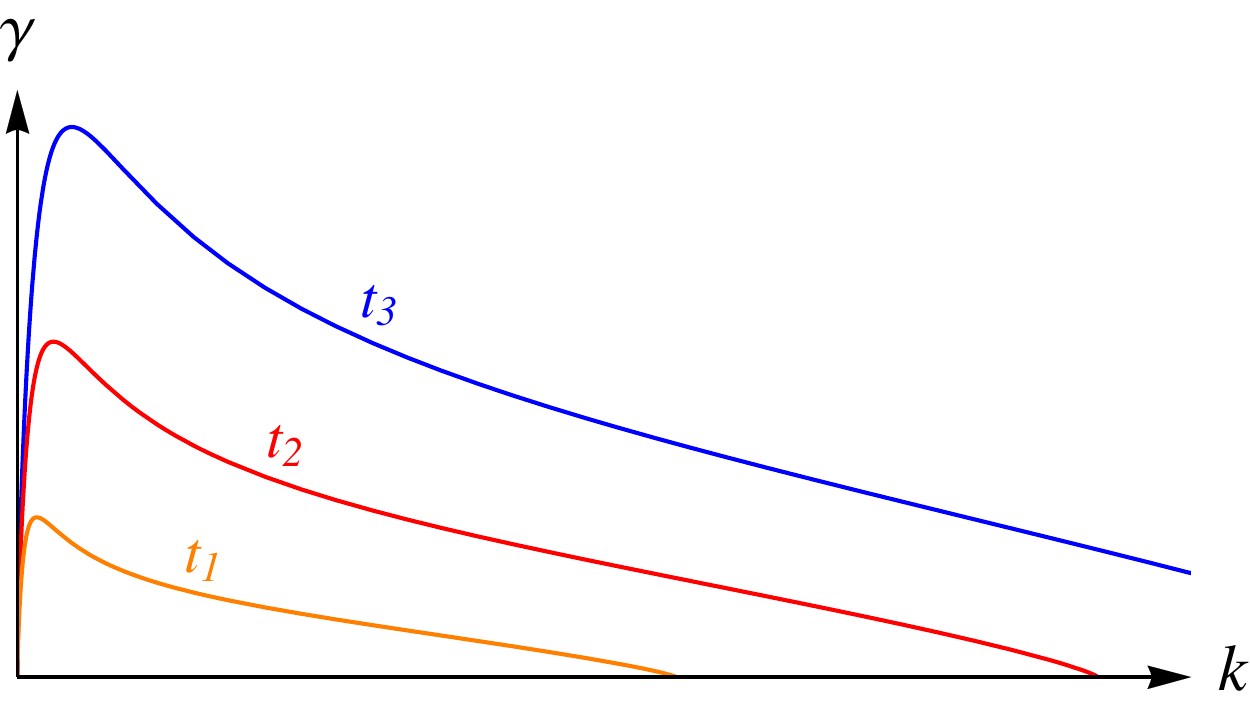}
\end{center}
\caption{Sketch of a typical tearing mode dispersion relation for a time-evolving current sheet, assuming that the current sheet is thinning in time ($t_1 < t_2 < t_3$).}
\label{fig02}
\end{figure}

In this Section, we provide a general framework to evaluate the properties of the plasmoid instability in general current sheets that can evolve over time. In such general current sheets, tearing modes \citep{Bisk00,Goedbloed2010,Fitz2014} do not begin to grow at the same time, i.e. they are rendered unstable at different times. Moreover, their growth rate does not depend solely on the wavenumber $k$, but also on the time $t$ that has elapsed since the current sheet evolution commenced at some initial aspect ratio (see example in Fig. \ref{fig02}).

The amplitude of the tearing modes evolves as per
\begin{equation} \label{PsiEvol}
{\psi}(k,t) = {\psi_{0}} \exp \Bigg(\int_{t_0}^t {{\gamma}(k,t')\,dt'}\Bigg) \, ,
\end{equation}
where $\gamma (k,t)$ indicates the time-dependent growth rate, $t_0$ is the initial time and $\psi_0 := \psi\left(k,t_0\right)$ represents the perturbation from which the modes can start to grow. If $\gamma$ were constant, the amplitude evolution would be identical to that obtained from conventional linear eigenmode theory. Since $\gamma$ itself depends on time, it is more instructive to regard (\ref{PsiEvol}) as the WKB solution \citep{BenOr78} to the linearized equations governing the tearing mode process. Notice that in the linear phase, the amplitudes of the modes are `small', and therefore they do not affect the current sheet evolution and there is no mode-mode coupling. Note also that we neglect mode stretching due to the reconnection outflow, which results to be a good approximation for the very large $S$ plasmas of interest in this work. We point out, however, that mode stretching is important to correctly evaluate the critical Lundquist number $S_c$ above which the plasmoid instability is manifested \citep[see][]{Huang2017}.

The linear phase is terminated when the plasmoid half-width ${w}(k,t)$ is on the same order as the inner layer width ${\delta_{{\rm{in}}}}(k,t)$. The former is given by \citep[see, for example,][]{Fitz2014}
\begin{equation} \label{half_width}
{w}(k,t) =  2{\left( {\dfrac{\psi\, a} B} \right)^{1/2}} \, ,
\end{equation}
where $\psi$, defined in (\ref{PsiEvol}), must be understood as being evaluated at the resonant surface. Here, $B$ represents the reconnecting magnetic field evaluated upstream of the current sheet, while $a$ is the half-width of the current sheet. The latter, namely ${\delta_{{\rm{in}}}}(k,t)$, is not the same for all physics models, and hence we shall leave it unspecified at this stage. 

As described above, we identify the end of the linear phase with the condition $w = \delta_{{\rm{in}}}$. Of course, this condition is not a sharp cutoff for the end of the linear phase. In reality, there could be a $\calo(1)$ factor that needs to be included prior to the equality sign. However, for the purposes of simplicity, we shall terminate the linear phase when these two quantities are exactly equal to one another. Finally, a caveat regarding $w$ must be introduced - it represents the plasmoid half-width only when its associated mode is much more dominant than the rest. This assumption can be slightly relaxed to cases where the perturbation amplitude is sufficiently localized in the spectrum. It turns out that this condition is typically met at the end of the linear stage of the plasmoid instability. A more precise evaluation of the plasmoid width can still be obtained by considering the contribution of a proper range of the fluctuation spectrum \citep{Huang2017}.

Although we have now specified the end of the linear phase, we have not still identified the tearing mode that emerges dominant at the end. At this stage, we introduce the primary physical principle behind the paper. We follow the approach espoused in \citet{CLHB16}, namely, the mode that emerges ``first'' at the end of the linear phase is the one that has taken the \emph{least time} to traverse it.  This ``principle of least time for the plasmoid instability'' shares some apparent similarities with the renowned Fermat's principle of least time, but there is also one essential difference - the latter relies upon a variational principle \citep{BoWo80}, whereas in the former the extremum of a function (the time) is computed.

\begin{figure}
\begin{center}
\includegraphics[width=8.6cm]{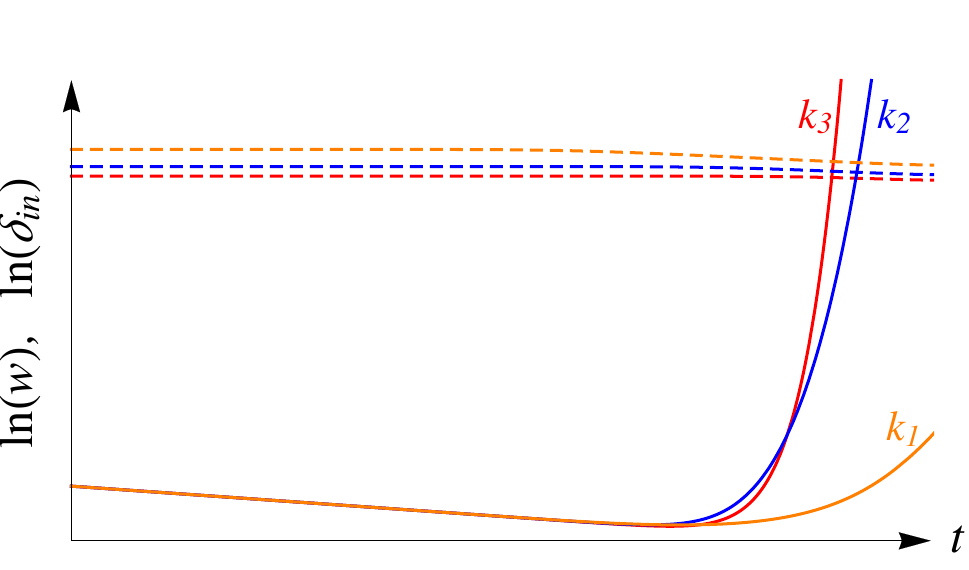}
\end{center}
\caption{Sketch of the typical dynamical evolution of three different modes starting from the same $w_0$ value, assuming an exponentially thinning current sheet. The solid lines represent the amplitude $\ln{w(k,t)}$ for three different modes with wavenumbers $k_1$ (orange), $k_2$ (blue), and $k_3$ (red). The dashed lines indicate the value $\ln{\delta_{{\rm{in}}}(k,t)}$ for wavenumbers $k_1$ (orange), $k_2$ (blue), and $k_3$ (red). The linear phase of the instability ends when the first wavenumber satisfies the condition $w = \delta_{\rm{in}}$, which corresponds to $k_3$ in this sketch. Note that while $k_2$ initially grows faster than $k_3$, at a later stage it is $k_3$ that dominates. It is also important to recognize that modes can be quiescent for a significant period of time before starting to grow when $\gamma (k,t) > 1/\tau$, with $\tau$ being the characteristic timescale of the current sheet evolution. Finally, observe that $w$ initially decreases because of the thinning of the current sheet.}
\label{fig03}
\end{figure}

\begin{figure}
\begin{center}
\includegraphics[width=8.6cm]{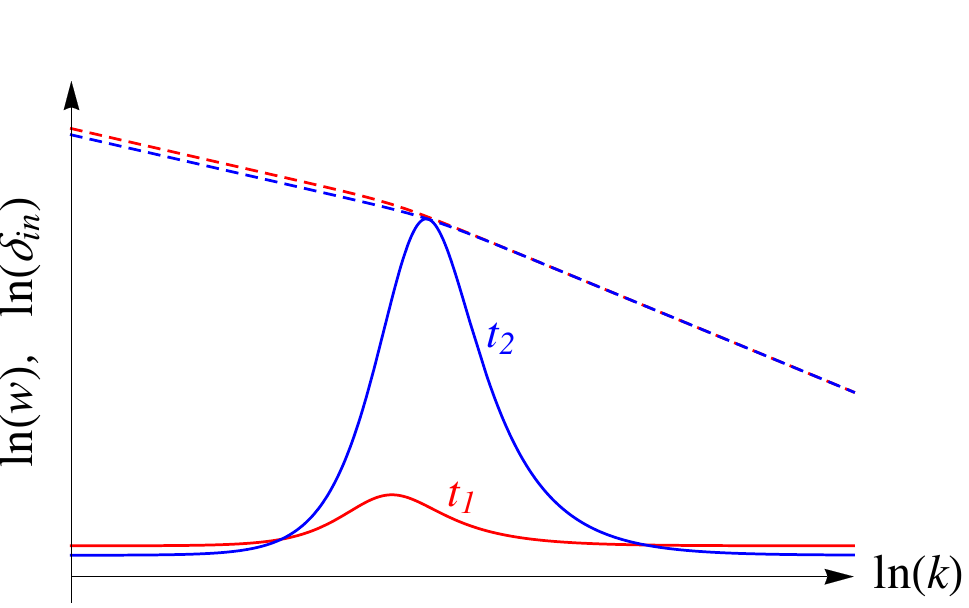}
\end{center}
\caption{Sketch of the typical spectrum of $\ln{w}$ and $\ln{\delta_{\rm{in}}}$ at two different times, with $t_1 < t_2$. The solid and dashed lines represent $\ln{w}$ and $\ln{\delta_{\rm{in}}}$, respectively, at time $t_1$ (red) and $t_2$ (blue). 
The linear phase of the instability ends when the first wavenumber satisfies the condition $w = \delta_{\rm{in}}$, which occurs at $t=t_2$ in this sketch. Observe that at $t=t_2$, the amplitude $\ln{w}$ for small and large wavenumbers decreases with respect to the time $t=t_1$ because of the current sheet thinning rate exceeding the growth rate for those wavelengths.}
\label{fig04}
\end{figure}

Some modes may become unstable from an early stage and continue growing at a steady (relatively slow) pace. Others may remain stable for a long time, therefore remaining quiescent, until they become unstable at a later stage and are subject to explosive growth (see example in Fig. \ref{fig03}). Thus, amongst this wide range of possibilities, the above principle enables us to select the mode that exits the linear stage first. In mathematical terms, these conditions are expressible as follows. Firstly, we have 
\begin{equation} \label{Cond1}
w\left(k_*,t_*\right) = \delta_{\mathrm{in}}\left(k_*,t_*\right) \, ,
\end{equation}
where the symbol `$*$' denotes the end of the linear phase. The above expression implies that the time $t_*$ is solely a function of the wavenumber $k_*$. Then, the principle of least time amounts to stating that 
\begin{equation} \label{Cond2LT}
\frac{dt_*}{dk_*} = 0 \, .
\end{equation}
Hence, the conditions (\ref{Cond1}) and (\ref{Cond2LT}) permit to compute the mode that takes the least value of $t_*$. 
It can also be shown \emph{a posteriori} that, in the neighborhood of $k_*$, $w$ is localized and has a stronger dependence than $\delta_{{\rm{in}}}$ on $k$. Therefore, the mode that completes the linear phase in the least time can be seen as the dominant one that enters the nonlinear phase (see example in Fig. \ref{fig04}).

We can explicitly rewrite Eq. (\ref{Cond1}) as
\begin{equation} \label{Eq_1a}
{\left. {\left\{ {\ln \left( {\frac{\delta_{\rm{in}}}{w_0}\frac{g^{1/2}}{f^{1/2}}} \right) - \frac{1}{2}\int_{t_0}^{t} {\gamma ({t}')d{t}'} } \right\}} \right|_{{k_*},{t_*}}} = 0 \, ,
\end{equation}
where ${w_0}: = {w}({k},{t}_0) = 2 \left({\psi_0}{a_0}/{B_0}\right)^{1/2}$ is the label for the initial perturbation amplitude and the functions $f$ and $g$ are defined such that $a(t)=a_0 f(t)$ and $B(t)=B_0 g(t)$. We can also rewrite Eq. (\ref{Cond2LT}) as 
\begin{equation} \label{Eq_1b}
{\left. {\frac{{\partial F(k,t)}}{{\partial k}}} \right|_{{k_*},{t_*}}} = 0
\end{equation}
if ${\left. {\partial /\partial t[F(k,t)]} \right|_{{k_*},{t_*}}} \ne 0$. Here we have defined 
\begin{equation} \label{Cond2}
F(k,t):= \delta_{{\rm{in}}}(k,t) - w(k,t) 
\end{equation}
as in our previous Letter \citep{CLHB16}. 

Hitherto, our discussion has been completely general as the above relations (\ref{Eq_1a}) and (\ref{Eq_1b}) describing the principle of least time are equally applicable for a wide range of plasma models that can include resistive, viscous and collisionless contributions. In this paper, we shall focus primarily on the former duo, i.e. the resistive and visco-resistive regimes.

\section{Resistive Regime} \label{SecResistive}

In what follows, we move to dimensionless quantities. We adopt a normalization convention such that all the lengths are normalized to the current sheet half-length $L$, the time to the Alfv{\'e}n time $\tau_A = L/v_A$, and the magnetic field to the upstream field ${B_0}$. Thus, the other quantities are normalized as
\begin{eqnarray} \label{normalization}
\left( {\hat k}, {\hat \gamma}, {\hat \psi}, {\hat \eta}, {\hat \nu} \right) = \left( kL, {\gamma {\tau_A}, \frac{\psi}{L{B_0}}, \frac{\eta}{L{v_A}}, \frac{\nu}{L{v_A}}} \right) \, ,
\end{eqnarray}
where we have used carets for denoting dimensionless quantities. Note that the normalized magnetic diffusivity corresponds to the inverse of the Lundquist number, i.e. ${{\hat \eta}^{-1}} = S$, while the normalized kinematic viscosity corresponds to the inverse of the kinetic Reynolds number when the Alfv{\'e}n velocity is the typical velocity scale of the system. Therefore, the ratio ${\hat \nu}/{\hat \eta} =: P_m$ defines the magnetic Prandtl number. 
\footnote{We specify that the magnetic Prandtl number that appears in the formulas of this paper is defined using the cross-field collisional value of $\hat \nu$.}

In this Section we consider the plasmoid instability in the resistive regime, which is characterized by $P_m \ll 1$, while the next section is devoted to the visco-resistive regime, in which $P_m \gg 1$.

Although the framework provided in Sec. \ref{SecLeastTime} is fully general, here we are interested in the case where the current sheet half-length $L$ and the reconnecting magnetic field $B$ remain approximately constant, while the current sheet width decreases in time. This is indeed a classic case of current sheet formation (see also \citealt{Huang2017}).
The function $f(\hat t) = {\hat a}(\hat t)/{\hat a}_0$ that takes into account the current sheet thinning must obey $f({{\hat t}_0}) = 1$ and 
\begin{equation} \label{}
\mathop {\lim }\limits_{\hat t \to \infty } f(\hat t) = \frac{{(1 + {P_m})}^{1/4}}{{{\hat a}_0}S^{1/2}} \, .
\end{equation} 
Indeed, $\hat a = S^{-1/2} (1+P_m)^{1/4}$ is the natural lower limit to the thickness of a reconnection layer \citep{Sweet1958,Parker1957,Park1984}. In the resistive regime, we have simply ${\lim_{\hat t \to \infty }}\,f(\hat t) = \hat a_0^{-1}{S^{- 1/2}}$.

When the modes grow slower than the evolution of the current sheet, i.e. ${\hat \gamma} < {{\hat a}^{ - 1}}d\hat a/d\hat t$, the change in ${\hat w}$ is dominated by the change in ${\hat a}$ and the growth rate ${\hat \gamma}$ is negligible in this respect. On the other hand, when the modes grow faster than the evolution of the current sheet, i.e. ${\hat \gamma} > {{\hat a}^{ - 1}}d\hat a/d\hat t$, the change in ${\hat w}$ is mainly due to the growth rate of the perturbed magnetic flux. In this case, the tearing modes growth rate can be computed using the instantaneous value of ${\hat a}$ in the standard tearing mode dispersion relations. Depending on the value of the tearing stability parameter ${{\hat \Delta}'}$ \citep{FKR_1963}, two simple algebraic relations can be considered. For ${{\hat \Delta}'} {{\hat \delta }_{{\rm{in}}}} \ll 1$ (the small-${{\hat \Delta}'}$ regime), tearing modes grow as per the relation \citep{FKR_1963} 
\begin{equation} \label{FKRDisp}
\hat{\gamma}_s \simeq c_{\Gamma} \hat{k}^{2/5} \hat{a}^{-2/5} {\hat \Delta '^{4/5}} S^{-3/5} \, ,
\end{equation} 
where $c_{\Gamma}  = {\big[ {{{(2\pi )}^{ - 1}}\Gamma (1/4)/\Gamma (3/4)} \big]^{4/5}} \approx 0.55$. On the other hand, for ${{\hat \Delta}'} {{\hat \delta }_{{\rm{in}}}} \gtrsim 1$ (the large-${{\hat \Delta}'}$ regime), the growth rate becomes \citep{Coppi1976, Ara1978} 
\begin{equation} \label{CoppiDisp}
\hat{\gamma}_l \simeq \hat{k}^{2/3} \hat{a}^{-2/3} S^{-1/3} \, .
\end{equation}
In our analysis, we are interested in the complete ${{\hat \Delta}'}$-domain. Therefore, we seek an expression for $\hat{\gamma}$ that (i) is a reasonable approximation of the exact dispersion relation  \citep{Coppi1976, Ara1978}, (ii) reduces to (\ref{FKRDisp}) and (\ref{CoppiDisp}) in the proper limits, and (iii) is simple enough to be analytically tractable. For this purpose we adopt the half-harmonic mean of this two relations, namely
\begin{equation} \label{fullgamma}
\hat \gamma  = {{\hat \gamma }_s}\,{{\hat \gamma }_l}/({{\hat \gamma }_s} + {{\hat \gamma }_l}) \, ,
\end{equation}
which fulfills all of the criteria described above. If a different approximation for $\hat \gamma$ is adopted, such as the simpler one employed in several previous works \citep{TS97,BHYR,LSU13,HB13,PV14}, or the more complex one used by \citet{Huang2017},
the same scaling relations as the ones derived below are obtained, albeit with slightly different numerical factors.

At this point we only need to specify the inner resistive layer width, which corresponds to \citep{Fitz2014}
\begin{equation}
{{\hat \delta }_{{\rm{in}}}} = {{\hat \gamma }^{1/4}}{(\hat a/\hat k)^{1/2}}{S^{ - 1/4}} \, .
\end{equation}
Using this expression, Eqs. (\ref{Eq_1a}) and (\ref{Eq_1b}) can be combined to obtain the {\it least time equation} 
\begin{equation} \label{least_time_RES}
{\left. {\left\{ {\left( {\hat \gamma \bar t - \frac{1}{2}} \right)\frac{{\partial \hat \gamma }}{{\partial \hat k}} + \frac{{\hat \gamma }}{{\hat k}} + \frac{{\hat \gamma }}{{{{\hat \psi }_0}}}\frac{{\partial {{\hat \psi }_0}}}{{\partial \hat k}}} \right\}} \right|_{{{\hat k}_*},{{\hat t}_*}}} = 0 \, ,
\end{equation}
where $\bar t = \partial /\partial \hat \gamma \int_{{{\hat t}_0}}^{\hat t} {\hat \gamma (\hat t') d\hat t'}$. 
In our subsequent discussion, we assume that the natural noise of the system has a general power law form, namely ${{\hat \psi}_0} = {\varepsilon} {{\hat k}^{- \alpha}}$, but other cases can be treated on an equal footing by considering different perturbation spectra.
We also assume that the current sheet has the common Harris-type structure \citep{Harris1962}, for which ${\hat \Delta}' = 2 \big[(\hat{k}\hat{a})^{-1} - \hat{k}\hat{a}\big]/ \hat{a}$. This expression for ${\hat \Delta}'$ can be simplified by considering only the regime $\hat k \hat a \ll 1$, since the very slow-growing part of the mode evolution does not affect the results of the linear phase. Then, from Eq. (\ref{least_time_RES}) we get
\begin{eqnarray} \label{least_time_RES2}
 && \frac{1}{{\bar t}_*}  \Big[  \Big( {5 - \frac{{15\alpha}}{2}} \Big) \big(1 + {{\hat k}_*}^{ - 16/15}{{\hat a}_*}^{ - 4/3}{S^{ - 4/15}} \big) \nonumber \\
 && \quad \quad \quad  +  \Big( {9 - \frac{{15\alpha}}{2}} \Big) \big(1 + {{\hat k}_*}^{16/15}{{\hat a}_*}^{4/3}{S^{4/15}} \big) \Big]   \nonumber\\
 &&\qquad= 3{{\hat k}_*}^{2/3}{{\hat a}_*}^{-2/3}{S^{-1/3}} - 5{{\hat k}_*}^{-2/5}{{\hat a}_*}^{-2}{S^{-3/5}} \, .
\end{eqnarray} 
It can be shown {\it a posteriori} that the two terms on the right-hand-side must approximately balance each other for ${{\hat w} ({{\hat k}_*},{{\hat t}_0}) } \ll {{\hat k}_*}^{-3/5} S^{-2/5}$. Hence, the emergent mode satisfies the relation 
\begin{equation} \label{k_dominant}
{\hat k_ *} \simeq {c_k} {{\hat a}_*}^{ - 5/4}{S^{ - 1/4}} ,
\end{equation}
where ${c_k}$ is a $\mathcal{O}(1)$ coefficient. This implies that
\begin{equation} \label{gamma_dominant}
{{\hat \gamma }_ *} \simeq {c_\gamma} {{\hat a}_*}^{-3/2} S^{-1/2} \, ,
\end{equation}
\begin{equation} \label{delta_dominant}
{{\hat \delta }_{{\rm{in}}*}} \simeq {c_\delta} {{\hat a}_*}^{3/4} S^{-1/4} \, ,
\end{equation}
where ${c_\gamma}$ and ${c_\delta}=(c_\gamma/c_k^{2})^{1/4}$ are also $\mathcal{O}(1)$ coefficients. The above relations indicate that the dominant mode at the end of the linear phase has the same scaling properties of the fastest growing mode \citep{FKR_1963}, the latter of which satisfies the equation ${{\partial \hat \gamma /\partial \hat k} |_{{{\hat k}_{f*}},{{{\hat t}_*}}}} = 0$. 

For moderately high values of the Lundquist number $S$, the reconnecting current sheet might be capable of attaining the Sweet-Parker inverse-aspect-ratio ${{\hat a}_*} \simeq S^{-1/2}$. In this case, it is straightforward to see that
\begin{equation} \label{SP_scalings}
{{\hat \gamma }_*} \sim S^{1/4} \, , \quad {\hat k_*} \sim {S^{3/8}} \, , \quad {{\hat \delta }_{{\rm{in}}*}} \sim  S^{-5/8} \, .
\end{equation}
It is therefore not surprising to discover that these relations match the ones obtained in previous studies of the plasmoid instability \citep{TS97,LSC07,BHYR,BBH12,HB13,ComGra16} that were undertaken assuming a fixed Sweet-Parker current sheet.

On the other hand, for very high Lundquist numbers, which are widely prevalent in most astrophysical plasmas, the plasmoids complete their linear evolution well before the Sweet-Parker aspect ratio is reached. Thus, we need to calculate ${\hat a}_*$ for a more general case. This can be done by substituting the relations for the dominant mode at the end of the linear phase into Eq. (\ref{Eq_1a}), which yields the following equation for the inverse-aspect-ratio:
\begin{equation} \label{eq_3a} 
\ln \left( {\frac{{{c_\delta }}}{{2\hat \psi _0^{1/2}}}\frac{{\hat a_*^{1/4}}}{{{S^{1/4}}}}} \right) = \frac{1}{2}\int_{{{\hat a}_0}}^{{{\hat a}_*}} {\hat \gamma (\hat a)\frac{{d\hat t}}{{d\hat a}}d\hat a} \, .
\end{equation}
This equation gives us the final inverse-aspect-ratio ${\hat a}_*$ for a general current sheet evolution ${\hat a}({\hat t})$. It is evident that ${\hat a}_*$, and thus the scaling relations of ${{\hat \gamma }_*}$, ${\hat k_*}$, ${{\hat \delta }_{{\rm{in}}*}}$ and ${\hat t_*}$, cannot be universal, because they depend on the specific form of the function ${\hat a}({\hat t})$.

Since we must specify a specific form of ${\hat a}({\hat t})$, in what follows, we first consider what is probably the most typical case of current sheet thinning, namely the exponential thinning. This is indeed known to be standard case for instability-driven current sheets. Then, we generalize the results of the exponential thinning to include also algebraic cases. Other, less common, possibilities could also be investigated, since the developed framework is general.

\subsection{Exponentially shrinking current sheet}\label{SecExpCurr}

It can be shown from first principles that the exponential thinning of a reconnecting current sheet evolves according to the expression \citep{Kuls05}
\begin{equation} \label{ExpCurr}
\hat a{(\hat t)^2} = (\hat a_0^2 - \hat a_\infty ^2){e^{ - 2\hat t/\tau }} + \hat a_\infty ^2  \, ,
\end{equation}
where $\hat a_\infty = S^{-1/2}$ in the resistive regime \citep{Sweet1958,Parker1957}. This expression slightly differs from the one we adopted in our previous work \citep{CLHB16}, but it shares the same asymptotic behaviors for small and large $\hat t$, therefore leading to the same asymptotic relations for the plasmoid instability. Other cases for $\hat a_\infty$, such as the ones imposed in the numerical simulations by \citet{Tenerani2015}, where $\hat a_\infty = S^{-1/3}$, are not considered here, because they are not supported by physical evidence \citep{Kuls05,Huang2017}.
In this respect, it is worth noting that the plasmoid half-width (\ref{half_width}) starts to grow only when ${\hat \gamma } > 1/\tau$ (for ${\hat \gamma } < 1/\tau$ it is straightforward to check that ${\hat w}({\hat t})$ decreases because of the rapid decrease of ${\hat a}({\hat t})$). We will see later that this condition occurs when ${\hat a} < {{\hat k}_*} {S^{ - 1/2}}{\tau ^{3/2}}$, which is smaller than $S^{-1/3}$ for $\tau$ of order unity, which is indeed the case for an ideal exponentially thinning current sheet \citep{Kuls05,Huang2017}.

Using Eq. (\ref{ExpCurr}) we can compute the {\it transitional} time ${{\hat t}_T}$ that separates the two asymptotic behaviors for small and large $\hat t$. For ${\hat t} > {{\hat t}_T}$, with 
\begin{equation}\label{threshold1}
{{\hat t}_T} = \tau \ln \sqrt {1 + \hat a_0^2/\hat a_\infty ^2} \, , 
\end{equation}
we have ${{\hat a}_*} \simeq \hat a_\infty$. On the other hand, for ${\hat t} < {{\hat t}_T}$ we have $\hat a = \hat a_0 {e^{- \hat t/\tau }}$. While for ${\hat t} > {{\hat t}_T}$ one recovers the relations (\ref{SP_scalings}), the case ${\hat t} < {{\hat t}_T}$, which is more relevant for astrophysical environments because it occurs for larger $S$-values, necessitates further analysis. In this case we have to solve Eq. (\ref{eq_3a}). Using ${\hat a}_* \ll \hat{a}_0$, we obtain
\begin{equation} \label{impl_res_a} 
{{\hat a}_*}\, \simeq {6^{2/3}}{c_a}\frac{{{\tau ^{2/3}}}}{{{S^{1/3}}}}{\left[ {\ln \left( {\frac{{c_\delta ^6c_k^{3\alpha }}}{{{2^6}{\varepsilon ^3}}}\frac{{\hat a_*^{3(2 - 5\alpha )/4}}}{{{S^{3(2 + \alpha )/4}}}}} \right)} \right]^{ - 2/3}} \, ,
\end{equation} 
where 
\begin{equation}
{c_a} = {\left( {\frac{3}{4}{{\tilde c}_k}c_k^{2/3}} \right)^{2/3}}
\end{equation} 
\begin{equation}
{{\tilde c}_k} = 1 + c_k^{8/15}\left( {{{\cot }^{ - 1}}(1 + \sqrt 2 c_k^{4/15}) + {{\cot }^{ - 1}}(1 - \sqrt 2 c_k^{4/15})} \right) .
\end{equation} 
The coefficient $c_a$ turns out to be $c_a \approx 0.3$ for $c_k$ of order unity. Therefore, we can neglect the factor $6^{2/3} c_a \approx 1$ in Eq. (\ref{impl_res_a}). This equation for the inverse-aspect-ratio can be solved exactly in terms of the Lambert $W$ function \citep{Corless1996}, but here we prefer to consider an asymptotic solution that yields more transparent results. As was done in \citet{CLHB16}, we solve Eq. (\ref{impl_res_a}) by iteration obtaining
\begin{equation}\label{espl_res_a_GENERAL} 
{{\hat a}_*} \simeq \frac{{{\tau ^{2/3}}}}{{{S^{1/3}}}}{\left( {\ln {\theta _R}} \right)^{ - 2/3}} \, ,
\end{equation}
where
\begin{equation}
\theta_R := \frac{{{\tau ^{(2 - 5\alpha )/2}}}}{{{2^6}{\varepsilon ^3}{S^{(4 - \alpha )/2}}}} \, ,
\end{equation}
and the subdominant term proportional to $\ln \big(c_\delta^6 c_k^{3\alpha} \big)$ has been neglected.


Given the final inverse-aspect-ratio, we can easily determine the growth rate, wavenumber and inner layer width at the end of the linear phase:
\begin{equation} \label{espl_res_gamma_GENERAL} 
{{\hat \gamma }_*} \simeq  c_\gamma \frac{\ln {\theta_R}}{\tau} \, ,
\end{equation}
\begin{equation}\label{espl_res_k_GENERAL} 
{{\hat k}_*} \simeq c_k S^{1/6} {\left( {\frac{{\ln {\theta _R}}}{\tau }} \right)^{5/6}} \, ,
\end{equation}
\begin{equation}\label{espl_res_delta_GENERAL} 
{{\hat \delta }_{{\rm{in*}}}} \simeq \frac{c_\delta}{S^{1/2}}  {\left( {\frac{\tau }{{\ln {\theta _R}}}} \right)^{1/2}} \, .
\end{equation}
These relations exhibit a non-trivial dependence with respect to the Lundquist number $S$, the noise level $\hat \psi_0$ (through both $\varepsilon$ and $\alpha$), and the timescale of the driving process $\tau$.  Note that these scaling relations of the plasmoid instability are not pure power laws, as they also include \emph{non-negligible logarithmic factors}. This has important implications for very large-$S$ plasmas like those typically encountered in astrophysical environments \citep{JD11}, since the scaling properties of the plasmoid instability change considerably with respect to those obtained for not-so-large $S$ plasmas.

To better evaluate the implications of the new scaling relations, let us focus on the case in which the natural noise amplitude is approximately the same for all wavelengths. In this case we can set $\alpha = 0$, and, recalling that ${\hat w}_0 = 2 ({{\varepsilon} \, {\hat a_0}})^{1/2}$, Eqs. (\ref{espl_res_a_GENERAL})-(\ref{espl_res_delta_GENERAL}) reduce to 
\begin{equation}\label{espl_res_a2} 
{{\hat a}_*} \simeq \frac{\tau^{2/3}}{S^{1/3}} {\left[ \ln \left( {\frac{\tau}{S^2}\frac{\hat a_0^3}{{\hat w}_0^6}} \right) \right]^{ - 2/3}} \, ,
\end{equation}
\begin{equation}\label{espl_res_gamma2} 
{{\hat \gamma }_*} \simeq  \frac{c_\gamma}{\tau } \, \ln \left( {\frac{\tau}{S^2}\frac{\hat a_0^3}{{\hat w}_0^6}} \right) \, ,
\end{equation}
\begin{equation}\label{espl_res_k2} 
{{\hat k}_*} \simeq c_k \frac{S^{1/6}}{{\tau ^{5/6}}}{\left[ {\ln \left( {\frac{{\tau}}{S^2}\frac{\hat a_0^3}{{\hat w}_0^6}} \right)} \right]^{5/6}} \, ,
\end{equation}
\begin{equation}\label{espl_res_delta2} 
{{\hat \delta }_{{\rm{in*}}}} \simeq {c_\delta } {\left( {\frac{\tau }{S}} \right)^{1/2}}{\left[ {\ln \left( {\frac{\tau}{S^2}\frac{{\hat a_0^3}}{{{{\hat w}_0^6}}}} \right)} \right]^{ - 1/2}} \, .
\end{equation}
Note that these expressions are identical to those obtained in \citet{CLHB16} 
once that the quantity $6^{2/3} c_a$ is not explicitly set to unity.

\begin{figure}
\begin{center}
\includegraphics[width=8.6cm]{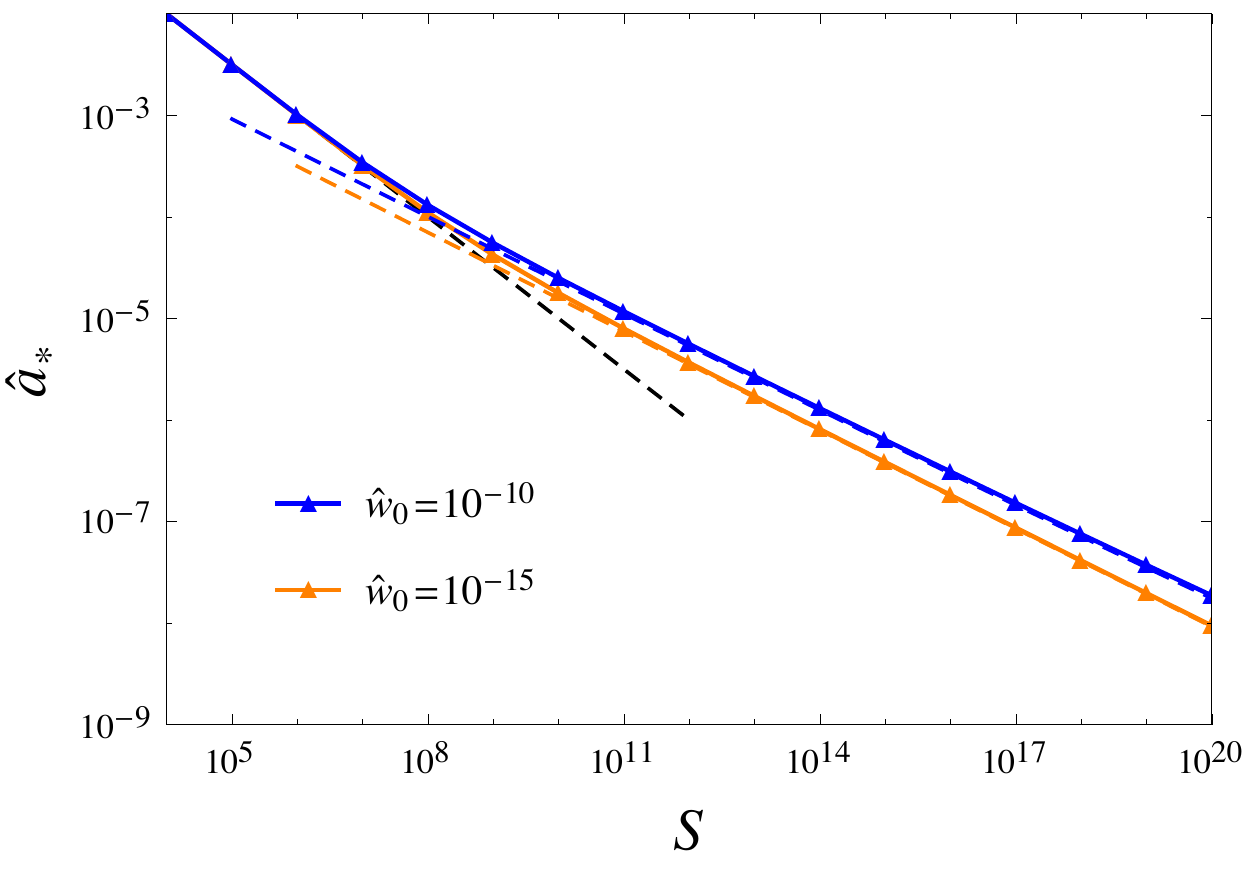}
\end{center}
\caption{Final inverse-aspect-ratio ${{\hat a}_*}$ as a function of the Lundquist number $S$ for ${\hat w}_0 = 2 ({{\varepsilon} \, {\hat a_0}})^{1/2} = 10^{-10}$ (blue) and ${\hat w}_0 = 2 ({{\varepsilon} \, {\hat a_0}})^{1/2} = 10^{-15}$ (orange). In both cases $\tau = 1$ and ${\hat a}_0 = 1/\pi$. The black dashed line denotes the Sweet-Parker scaling ${{\hat a}_*} \sim S^{-1/2}$. The colored solid and dashed lines refer to the numerical [Eqs. (\ref{Eq_1a}) and (\ref{Eq_1b})] and analytic [Eq. (\ref{espl_res_a2})] solutions, respectively.}
\label{fig1}
\end{figure}
From Eq. (\ref{espl_res_a2}), together with Eq. (\ref{ExpCurr}), we can see that the final inverse-aspect-ratio turns out to be bounded between ${S^{ - 1/2}} < {{\hat a}_*} < {\tau ^{2/3}}{S^{-1/3}}$. Eq. (\ref{espl_res_a2}) also indicates that ${{\hat a}_*}$ decreases for smaller perturbation amplitudes ${\hat \psi}_0$. 
The final inverse-aspect-ratio as a function of $S$ for two different values of ${\hat w}_0$ is plotted in Fig. \ref{fig1}. An inspection of this figure reveals that the Sweet-Parker aspect ratio can be attained only for moderately high $S$-values. The domain of existence of the Sweet-Parker aspect ratio may be slightly extended in lower noise systems, but, for most of the astrophysically relevant regimes, the final width of the reconnecting current sheet remains thicker as predicted by Eq. (\ref{espl_res_a2}).

\begin{figure}
\begin{center}
\includegraphics[width=8.6cm]{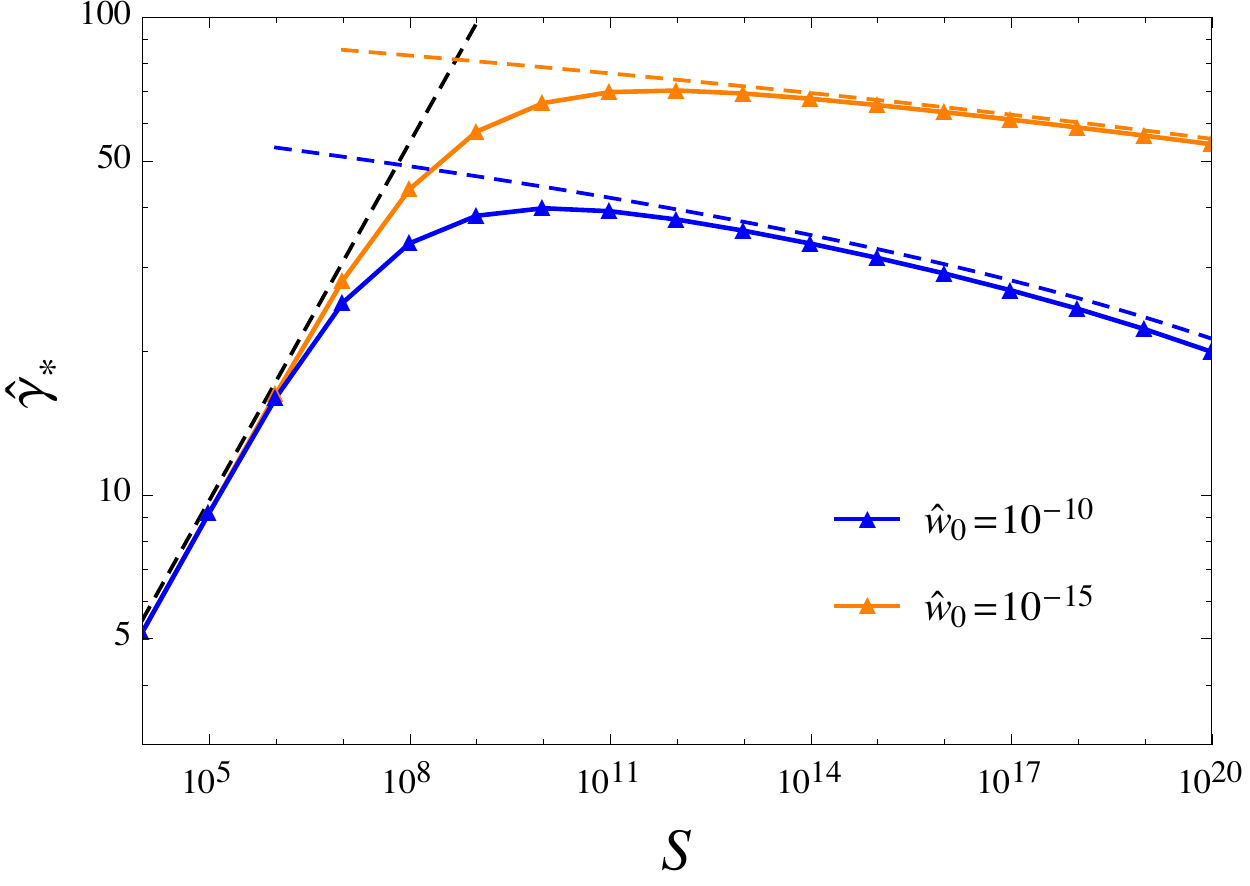}
\end{center}
\caption{Final (dominant mode) growth rate ${{\hat \gamma }_*}$ as a function of the Lundquist number $S$ for the same parameters adopted in Fig. \ref{fig1}. The black dashed line denotes the Sweet-Parker based scaling ${{\hat \gamma }_*} \sim S^{1/4}$. The colored solid and dashed lines refer to the numerical [from Eqs. (\ref{Eq_1a}) and (\ref{Eq_1b})] and analytic [Eq. (\ref{espl_res_gamma2})] solutions, respectively.}
\label{fig2}
\end{figure}

\begin{figure}
\begin{center}
\includegraphics[width=8.6cm]{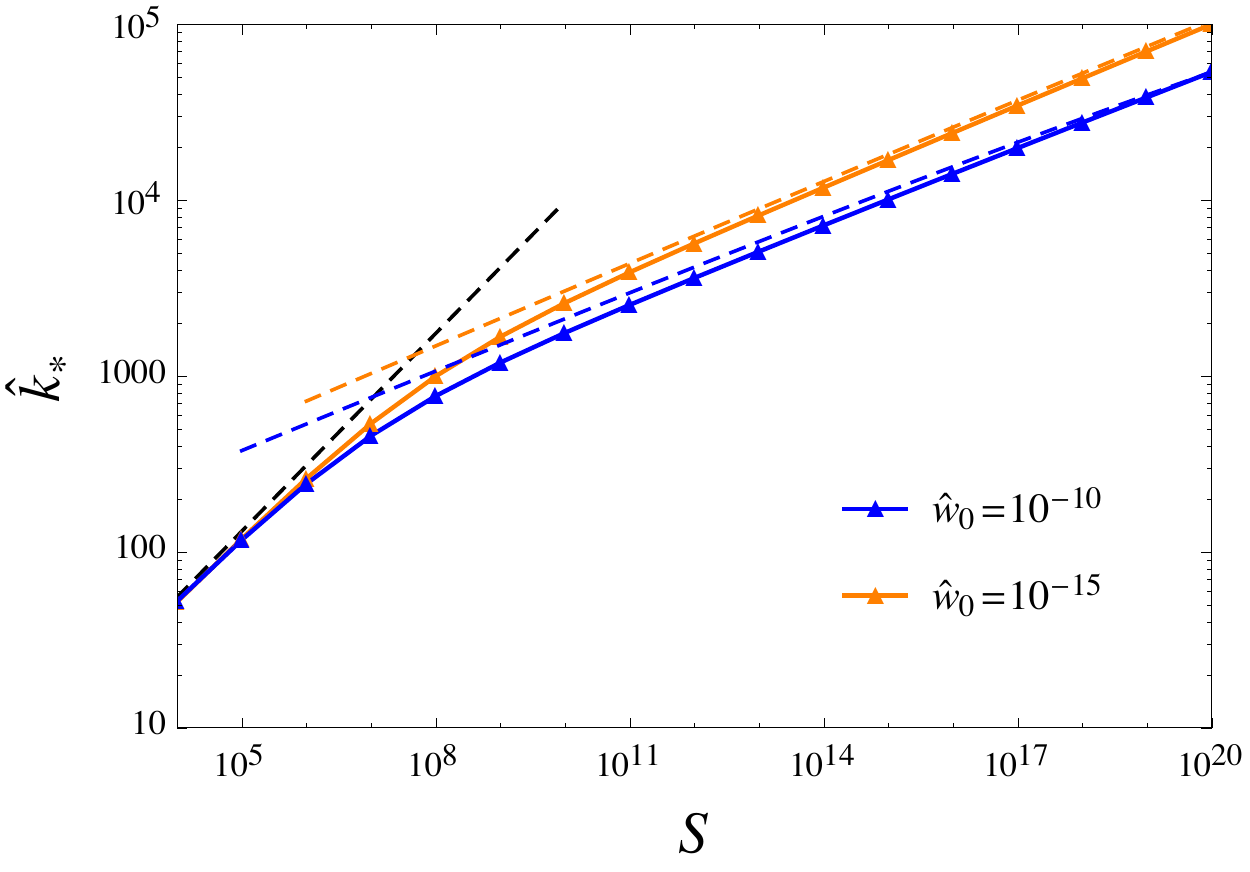}
\end{center}
\caption{Final (dominant mode) wavenumber ${{\hat k}_*}$ as a function of the Lundquist number $S$ for the same parameters adopted in Fig. \ref{fig1}. The black dashed line denotes the Sweet-Parker based scaling ${{\hat k}_*} \sim S^{3/8}$. The colored solid and dashed lines refer to the numerical [from Eqs. (\ref{Eq_1a}) and (\ref{Eq_1b})] and analytic [Eq. (\ref{espl_res_k2})] solutions, respectively.}
\label{fig3}
\end{figure}

The dependence of the growth rate ${{\hat \gamma}_*}$ and the wavenumber ${{\hat k}_*}$ as a function of the Lundquist number $S$ change significantly upon considering large $S$ systems. This is clearly shown in Figs. \ref{fig2} and \ref{fig3}, where the black dashed lines represent the earlier scalings, which are clearly not applicable to large-$S$ plasmas, while the solid curves represent the results that have been obtained by means of this new theoretical approach. The behavior of ${{\hat \gamma}_*}$ is non-monotonic in $S$, while ${{\hat k}_*}$ displays a monotonic behavior, but with much lower values with respect to the Sweet-Parker-based solution for large values of $S$. 
While counterintuitive at first glance, the decrease of the final growth rate for very large $S$ can be understood by noting that the inner layer width decreases for increasing $S$, therefore, a given noise amplitude leads to perturbation amplitudes closer to the condition for the end of the linear phase if $S$ is larger. This, in turn, reduces the time available for the acceleration of the perturbation growth.

Our approach also enables a quantification of the effects of noise: lower values of the noise can increase the final instantaneous growth rate and the number of plasmoids (which is proportional to ${{\hat k}_*}$), as can be seen from the orange curves in Figs. \ref{fig2} and \ref{fig3}. 
Finally, note that for large-$S$ astrophysical environments, Eq. (\ref{espl_res_delta2}) (not plotted here) indicates that the inner resistive layer width at the end of the linear phase is thicker than what would be predicted using the Sweet-Parker-based solution (\ref{SP_scalings}).

\begin{figure}
\begin{center}
\includegraphics[width=8.6cm]{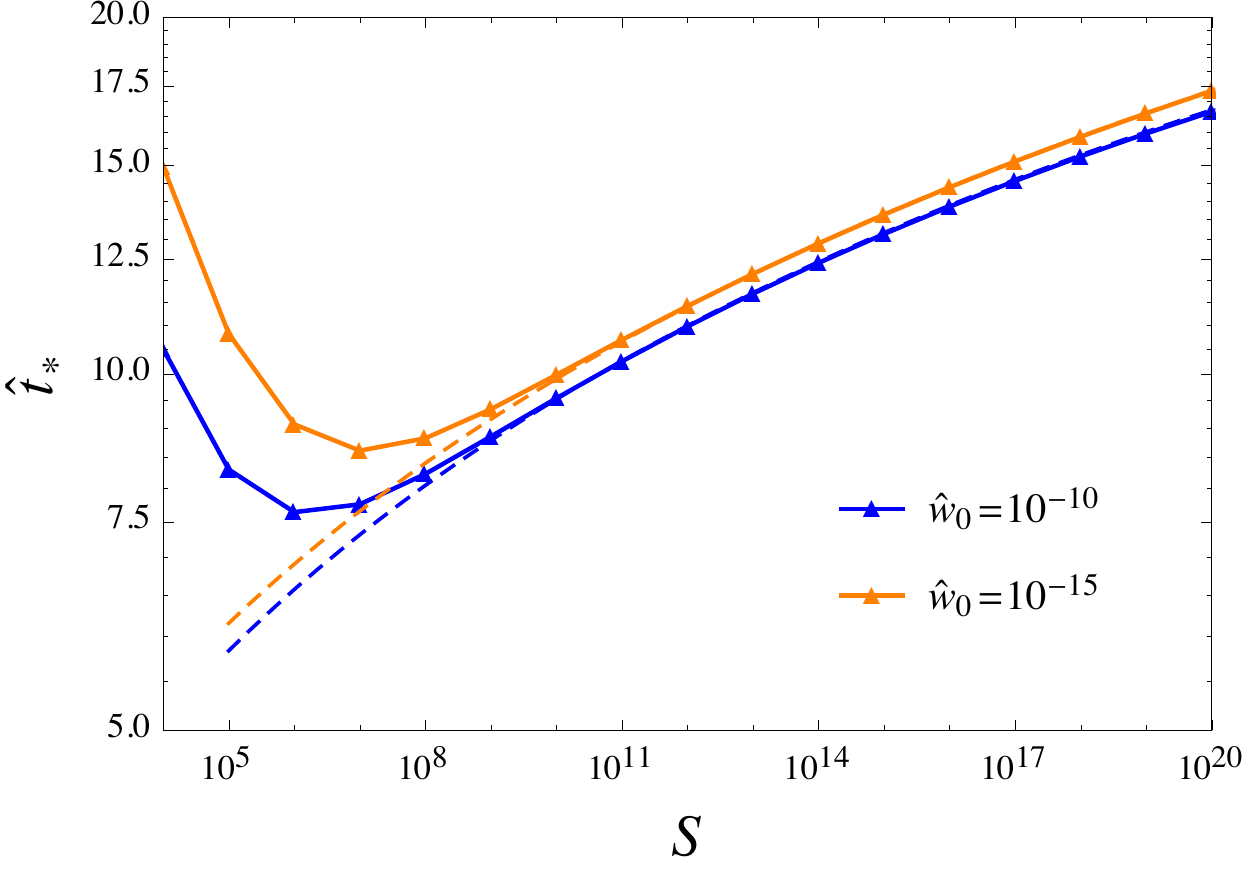}
\end{center}
\caption{Final time ${{\hat t}_*}$ as a function of the Lundquist number $S$ for the same parameters adopted in Fig. \ref{fig1}. The solid and dashed lines refer to the numerical [from Eqs. (\ref{Eq_1a}) and (\ref{Eq_1b})] and analytic [Eq. (\ref{fin_time})] solutions, respectively.}
\label{fig4}
\end{figure}
An important observable that can be duly obtained from Eq. (\ref{espl_res_a_GENERAL}) is the time that has elapsed since the current sheet evolution commenced at the initial inverse-aspect-ratio ${\hat a}_0$. This timescale corresponds to 
\begin{equation} \label{fin_time}
{{\hat t}_*} \simeq \tau \ln \left[ { {\hat a}_0 \frac{{{S^{1/3}}}}{{{\tau ^{2/3}}}}{{\left( {\ln {\theta _R}} \right)}^{2/3}}} \right] \, .
\end{equation}
Fig. \ref{fig4} shows that the elepsed time computed by means of the principle of least time has a non-monotonic behavior, confirmed also by recent numerical simulations \citep{Huang2017}, and after reaching a minimum value at moderate $S$-values, it increases as predicted by Eq. (\ref{fin_time}). Note that the time ${{\hat t}_*}$ does not correspond to the time required for the plasmoids to grow. This is due to the fact that the final (dominant mode) wavenumber ${{\hat k}_*}$ remains quiescent for a certain period of time before being subject to violent growth over a short timescale (as shown in the example in Fig. \ref{fig03}).

The actual time that it takes for the final wavenumber to undergo finite growth is ${\tau _p} = \tau \ln ({{\hat a}_{\rm{on}}}/{{\hat a}_*})$, where ${\hat a}_{\rm{on}}$ is the inverse-aspect-ratio at the onset time, i.e. when
$\hat \gamma ({{\hat k}_*},{{\hat t}_{\rm{on}}}) = 1/\tau$. Using Eq. (\ref{fullgamma}) and retaining the dominant terms, we find ${{\hat a}_{\rm{on}}} \simeq {{\hat k}_*} {S^{ - 1/2}}{\tau ^{3/2}}$. 
Therefore, the intrinsic timescale $\tau_p$ of the plasmoid instability becomes  
\begin{equation} \label{tau_p_RES}
{\tau _p} \simeq \tau \ln  \left[ { c_k {{ \left( {\ln {\theta_R}} \right)}^{3/2}}} \right] \, .
\end{equation}
This timescale exhibits a very weak dependence on the Lundquist number and the natural noise of the system, meaning that the intrinsic timescale of the plasmoid instability is nearly universal for exponentially thinning current sheets.

\begin{figure}
\begin{center}
\includegraphics[width=8.6cm]{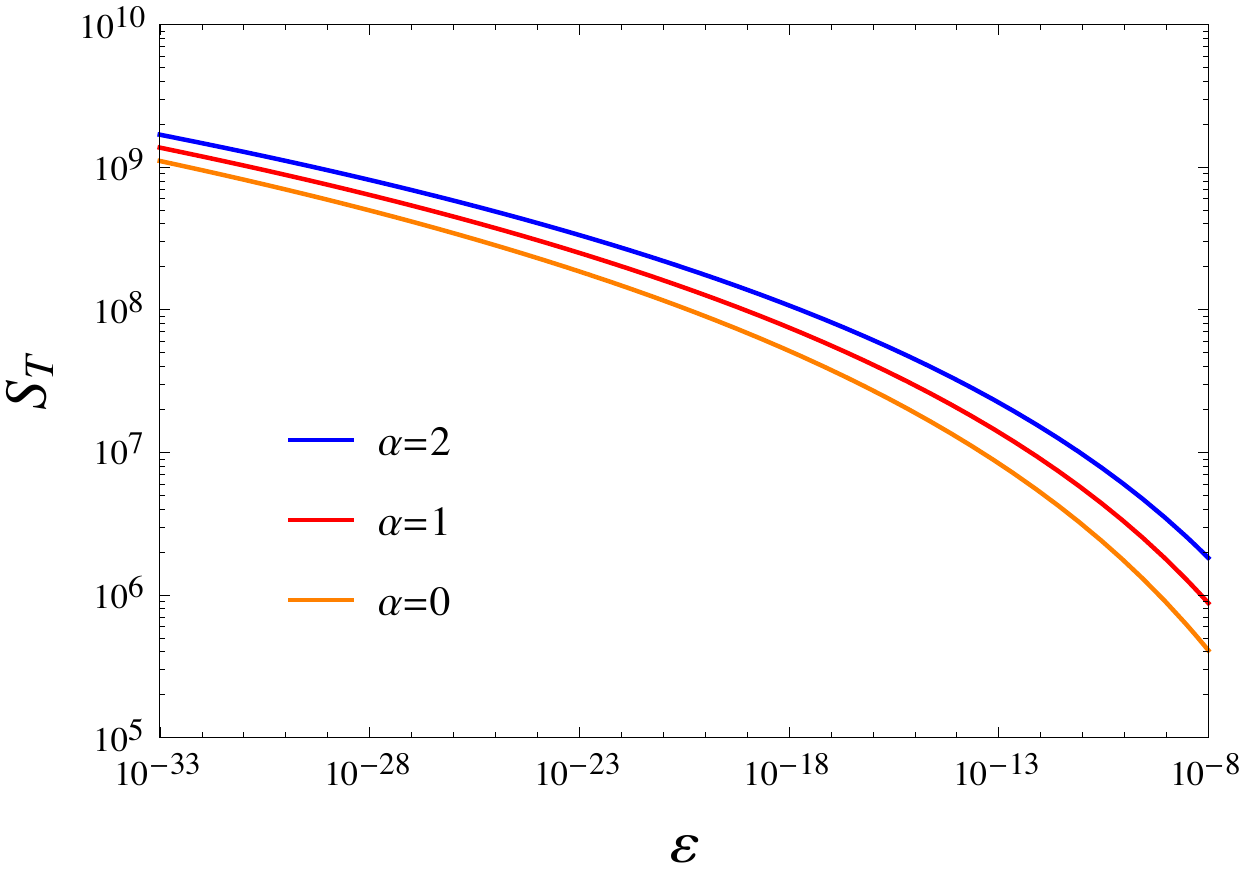}
\end{center}
\caption{Transitional Lundquist number $S_T$ as a function of the noise amplitude $\varepsilon$ for three different values of the spectral index $\alpha$. Recall that ${{\hat \psi}_0} = {\varepsilon} {{\hat k}^{- \alpha}}$. The curves are given by Eq. (\ref{ST_res_exact}) with $\tau=1$.}
\label{figS}
\end{figure}

Finally, we want to evaluate the value of the Lundquist number above which the scaling laws of the plasmoid instability change behavior as described by the previously obtained equations. We refer to this value as to the {\it transitional} Lundquist number. It can be obtained by equating the two asymptotic behaviors of ${\hat a}_*$, which yields the equation 
\begin{equation}
S_T^{-1/4} \, \ln \left( \frac{{{ \tau^{(2 - 5\alpha )/2}}}}{{2^6{\varepsilon ^3}{S_T^{(4 - \alpha )/2}}}} \right) = \tau \, .
\end{equation}
The exact explicit solution of this equation is 
\begin{equation}  \label{ST_res_exact} 
S_T = {\left[ {\frac{{\tilde \alpha }}{\tau }\,W\left( {\frac{1}{{\tilde \alpha }}{{\left( {\frac{{{ \tau ^{9(2-\alpha)/2}}}}{{2^6{\varepsilon^3}}}} \right)}^{1/\tilde \alpha }}} \right)} \right]^4}  \, ,
\end{equation}
where $\tilde \alpha :=  2(4-\alpha)$, and $W(z)$ is the Lambert $W$ function, which is defined such that $W(z) e^{W(z)} = z$.  
This expression exhibits a complex dependence on the noise level ($\varepsilon$ and $\alpha$), and the timescale of the driving process ($\tau$). A simpler asymptotic approximation can be constructed when considering large arguments of the Lambert $W$ function. In this case \citep{Corless1996}
\begin{equation}
W(z) = \ln (z) - \ln \big(\ln (z) \big) + o(1) \, .
\end{equation}
Keeping only the first term of this expansion, we obtain 
\begin{equation}
S_T = \frac{1}{\tau ^4}{\left[ {\ln \left(  {\frac{\tau^{9(2-\alpha)/2}}{ 2^6 \varepsilon^3  {\tilde \alpha }^{\tilde \alpha }}} \right)} \right]^4}  \, .
\end{equation}
From this expression we can see that $S_T$ decreases if the timescale of the current sheet thinning becomes larger. Furthermore, an increase of $S_T$ occurs for lower values of $\varepsilon$ and/or increasing values of $\alpha$. 
The accurate behavior of the transitional Lundquist number as a function of the system noise for a wide range of noise amplitudes is shown in Fig. \ref{figS}. The transitional Lundquist number turns out to be fairly modest even for very low noise amplitudes, implying that the plasmoid instability in most of the astrophysical systems should follow the newly obtained scaling laws.
%


\subsection{Generalized current sheet shrinking}\label{SecGenCurrShrink}

It is possible to generalize the results obtained for exponentially thinning current sheets in order to also describe current sheets whose thinning depends algebraically on time. While the former is the natural consequence of an instability-driven current sheet, the latter has been shown to occur in several cases of forced magnetic reconnection \citep{HK1985, WangBhatta1992, Fitz2003, Bhatta2004, Birn2005, Hossein2008, CGW15, ComissoJPP2015}. To encompass both exponential and algebraic behaviors, we consider a generalized current shrinking function of the form
\begin{equation}\label{GenCurrShrink}
\hat a{(\hat t)^2} = (\hat a_0^2 - \hat a_\infty ^2){\left( {\frac{\tau }{{\tau  + 2\hat t/\chi }}} \right)^\chi } + \hat a_\infty ^2 \, ,
\end{equation}
where $\hat a_\infty = S^{-1/2}$. This expression recovers the exponential thinning specified in Eq. (\ref{ExpCurr}) when taking the limit $\chi \to \infty$, while other cases can be obtained by considering different values of $\chi$. For example, setting $\chi = 2$, we obtain a current sheet thinning that is inversely proportional in time. This is relevant for various forced reconnection models, most notably the Taylor model \citep{HK1985, WangBhatta1992, Fitz2003, CGW15, ComissoJPP2015, Zhou2016, Beidler17}, which has applications in both laboratory and astrophysical plasmas.

For the adopted generalized current thinning function, the plasmoid half-width (\ref{half_width}) starts to grow when $\gamma  > {{\hat t}^{ - 1}}\ln {\left[ {(\tau  + 2\hat t/\chi )/\tau } \right]^{\chi /2}}$. Furthermore, the transitional time that separates the two asymptotic solutions for the plasmoid instability is 
\begin{equation}
{{\hat t}_T} = \frac{{\chi \tau }}{2}\left[ {{{\left( {1 + \frac{{\hat a_0^2}}{{\hat a_\infty ^2}}} \right)}^{1/\chi }} - 1} \right] \, .
\end{equation}
As before, we are especially interested in the case ${\hat t} < {{\hat t}_T}$, since it is astrophysically more relevant. However, to derive the analytical solution in this case, we consider only the small ${{\hat \Delta}'}$ branch of the dispersion relation in consideration of the fact that ${{\hat \gamma }_*} \approx {{\hat \gamma }_f}({{\hat t}_*})$, where ${{\hat \gamma }_f}$ is the instantaneous growth rate of the fastest growing mode. Thus, we approximate $\hat \gamma (\hat a)$ with ${{\hat \gamma }_s}(\hat a)$ in Eq. (\ref{eq_3a}), using again ${{\hat \Delta}'} {\hat a} \simeq 2(\hat{k}{\hat a})^{-1}$ and ${\hat a}_* \ll \hat{a}_0$. Therefore we obtain
\begin{equation}
\hat a_*^{1/\zeta} \, \ln \left( {\frac{{{c_\delta^6 }c_k^{3\alpha}}}{{2^6{{\varepsilon }^3}}}\frac{{\hat a_*^{3(2 - 5\alpha )/4}}}{{{S^{3(2 + \alpha )/4}}}}} \right) \sim  \hat a_0^{2/\chi }\frac{\tau }{{{S^{1/2}}}} \, ,
\end{equation} 
where 
\begin{equation}
\zeta : = \frac{{2\chi }}{{4 + 3\chi }} \, .
\end{equation}
Solving this equation along the same lines as Eq. (\ref{espl_res_a_GENERAL}), we obtain
\begin{equation}\label{GENespl_res_a4}
{{\hat a}_*} \sim {\left( {\frac{{\hat a_0^{2/\chi }\tau }}{{{S^{1/2}} \ln {\Theta_R}}}} \right)^\zeta } \, ,
\end{equation}
where
\begin{equation}
\Theta_R := \frac{{{{\left( {\hat a_0^{2\zeta /\chi }{\tau ^\zeta }{S^{ - \zeta /2}}} \right)}^{3(2 - 5\alpha )/4}}}}{{2^6{\varepsilon^3}{S^{3(2 + \alpha )/4}}}}  \, .
\end{equation}

From Eq. (\ref{GENespl_res_a4}), we obtain the following generalized scaling relations for the plasmoid instability:
\begin{equation}
{{\hat \gamma }_*} \sim {S^{(3\zeta  - 2)/4}}{\left( {\frac{{\ln {\Theta_R}}}{{\hat a_0^{2/\chi }\tau }}} \right)^{3\zeta /2}} \, ,
\end{equation}
\begin{equation}
{{\hat k }_*} \sim {S^{(5\zeta  - 2)/8}}{\left( {\frac{{\ln {\Theta_R}}}{{\hat a_0^{2/\chi }\tau }}} \right)^{5\zeta /4}} \, ,
\end{equation}
\begin{equation}\label{GENespl_res_delta4}
{{\hat \delta }_{{\rm{in*}}}} \sim {S^{ - (3\zeta  + 2)/8}}{\left( {\frac{{\hat a_0^{2/\chi }\tau }}{{\ln {\Theta_R}}}} \right)^{3\zeta /4}} \, .
\end{equation}
For $\chi \to \infty$, we recover the scaling relations (\ref{espl_res_a_GENERAL})-(\ref{espl_res_delta_GENERAL}), while different choices of $\chi$ give us the scaling relations relevant for different algebraic thinning possibilities. These expressions indicate that faster current sheet shrinking rates (larger $\chi$ and/or smaller $\tau$) lead to larger aspect ratio ($1/{{\hat a}_*}$), growth rate and wavenumber. On the other hand, the inner resistive layer width at the end of the linear phase decreases for faster current sheet formation.
It is also interesting to observe that for $\zeta > 2/5$ (i.e. $\chi > 2$), the number of plasmoids increases with $S$ in the astrophysically relevant regimes, but the opposite trend is possibly manifested for $\chi < 2$. In other words, for the latter case, the number of plasmoids can actually decrease in this regime as $S$ increases. For $\chi = 2$, where the thinning is inversely dependent on the time, the scaling of the number of plasmoids with $S$ is weak (logarithmic).

As it should be expected, the elapsed time from the initial aspect ratio can change significantly for different current sheet formation rates. Indeed, from Eqs. (\ref{GENespl_res_a4}) and (\ref{GenCurrShrink}), with ${\hat t} < {{\hat t}_T}$, the elapsed time results to be 
\begin{equation}
{{\hat t}_*} \sim  \frac{{\chi \tau }}{2}\left[ {\hat a_0^{(2\chi  - \zeta )/{\chi ^2}}{{\left( {\frac{{{S^{1/2}}}}{\tau }\ln {\Theta_R}} \right)}^{2\zeta /\chi }} - 1} \right] \, .
\end{equation}
Lower values of $\chi$ lead to much higher values of the elapsed time ${{\hat t}_*}$, implying that the final dominant wavenumber remains quiescent for a much longer period of time when the current sheet evolution is slower. 

The transitional Lundquist number $S_T$ for this class of generalized thinning current sheets can be computed by equating the two asymptotic branches for ${{\hat a}_*}$ in a manner analogous to that of exponential thinning. Thus, we are led to the equation
\begin{equation}
{S_T^{(\zeta  - 1)/2\zeta }} \,\ln {\Theta_{R_T}} \sim \hat a_0^{2/\chi }\tau  \, ,
\end{equation}
which can be inverted in a straightforward manner, by means of the Lambert $W$ function, to obtain $S_T$. 
Similarly, it is also possible to compute the timescale $\tau_p$ for the plasmoid instability in this generalized scenario by following the procedure outlined for exponential thinning sheets. 

We shall explore the implications of our preceding results for astrophysical plasmas in Section \ref{Discuss}. Next, we consider the visco-resistive regime and carry out a similar analysis.

\section{Visco-Resistive Regime} \label{SecViscoRes}

In this Section, we derive the corresponding scaling laws of the plasmoid instability in the presence of strong plasma viscosity, namely when $P_m \gg 1$. 
Plasma viscosity is indeed important in several astrophysical environments such as the (i) warm interstellar medium, (ii) protogalactic
plasmas, (iii) intergalactic medium and (iv) accretion discs around neutron stars and black holes \citep{KA1992,BraSub2005,BH08}. In this case ${\lim _{\hat t \to \infty }}f(\hat t) = \hat a_0^{-1} S^{-1/2} P_m^{1/4}$. Furthermore, two different relations for the growth rate in the small-${{\hat \Delta}'}$ and large-${{\hat \Delta}'}$ regimes must be considered. For ${{\hat \Delta}'} \hat{\delta}_{\rm{in}} \ll 1$, the growth rate of the tearing modes modified by strong plasma viscosity is \citep{BS1984} 
\begin{equation} \label{BSDisp}
\hat{\gamma}_s \simeq c_{1} \hat{k}^{1/3} \hat{a}^{-1/3} \hat{\Delta}'^{4/5} S^{-2/3} P_m^{-1/6} \, ,
\end{equation} 
where $c_1  = {({6^{2/3}}\pi)^{-1}}\Gamma (1/6)/\Gamma (5/6) \approx 0.48$. On the other hand, for ${{\hat \Delta}'} {{\hat \delta }_{{\rm{in}}}} \gtrsim 1$ the growth rate satisfies the relation \citep{Porcelli87}
\begin{equation} \label{PorcelliDisp}
{{\hat \gamma }_l} \simeq {c_{2}}{{\hat k}^{2/3}}{{\hat a}^{ - 2/3}}{S^{ - 1/3}}P_m^{ - 1/3} \, ,
\end{equation}
where $c_2 \approx 1.53$. An effective approximation for $\hat{\gamma}$ across the entire domain of ${{\hat \Delta}'}$ can be constructed as before, by using Eq. (\ref{fullgamma}).

In a manner analogous to the resistive regime, by combining Eqs. (\ref{Eq_1a}) and (\ref{Eq_1b}) and specifying the inner visco-resistive layer width \citep{Porcelli87}
\begin{equation}
{{\hat \delta }_{{\rm{in}}}} = {(\hat a/\hat k)^{1/3}}{S^{ - 1/3}}P_m^{1/6} \, ,
\end{equation} 
it is possible to obtain the {\it least time equation}
\begin{equation} \label{}
{\left. {\left\{ {\bar t \, \frac{{\partial \hat \gamma }}{{\partial \hat k}} + \frac{2}{{3\hat k}} + \frac{1}{{{{\hat \psi }_0}}}\frac{{\partial {{\hat \psi }_0}}}{{\partial \hat k}}} \right\}} \right|_{{{\hat k}_*},{{\hat t}_*}}} = 0 \, ,
\end{equation}
which follows from a careful application of Eq. (\ref{Cond2}).
By repeating the procedure delineated in the previous Section, we can find the counterpart of Eq. (\ref{least_time_RES2}) that is valid in the visco-resistive regime. This corresponds to 
\begin{eqnarray} \label{Fderdf2}
 && \frac{ 1 - \frac{3}{2}\alpha }{{\bar t}_*}  \Big[ \big(1 + c_2^{-1} {{\hat k}_*}^{-4/3}{{\hat a}_*}^{-5/3}{S^{-1/3}}{P_m^{1/6}} \big) \nonumber \\
 && \quad \quad  + \big(1 + {c_2} {{\hat k}_*}^{4/3}{{\hat a}_*}^{5/3}{S^{1/3}}{P_m^{-1/6}} \big) \Big]   \nonumber\\
 &&\qquad= {c_2}\hat k_*^{2/3}\hat a_*^{-2/3}{S^{-1/3}}P_m^{-1/3} \nonumber \\
 && \qquad \, - \hat k_*^{-2/3}\hat a_*^{-7/3}{S^{-2/3}}P_m^{-1/6} \, .
\end{eqnarray} 
It can be shown {\it a posteriori} that for ${{\hat w} ({{\hat k}_*},{{\hat t}_0}) } \ll {{\hat \delta }_{{\rm{in}}}} ({{\hat k}_*},{{\hat t}_0})$, the two terms on the right-hand-side must approximately balance each other. Thus, we end up with
\begin{equation} \label{k_dominant_VISCO}
{\hat k_ *} \simeq {\lambda_k} {{\hat a}_*}^{ - 5/4}S^{-1/4}P_m^{1/8} \, ,
\end{equation}
\begin{equation} \label{gamma_dominant_VISCO}
{{\hat \gamma }_ *} \simeq {\lambda_{\gamma}} {{\hat a}_*}^{-3/2} S^{-1/2} P_m^{-1/4} \, ,
\end{equation}
\begin{equation} \label{delta_dominant_VISCO}
{{\hat \delta }_{{\rm{in}}*}} \simeq {\lambda_{\delta}} {{\hat a}_*}^{3/4} S^{-1/4} P_m^{1/8} \, ,
\end{equation}
where ${\lambda_k}$, ${\lambda_\gamma}$ and ${\lambda_\delta}=\lambda_k^{-1/3}$ are $\mathcal{O}(1)$ coefficients. 

By assuming that the reconnecting current sheet has the time to reach the asymptotic value ${{\hat a}_*} \simeq S^{-1/2} P_m^{1/4}$, one can find
\begin{equation} \label{ViscoSP_scalings}
{{\hat \gamma }_*} \sim S^{1/4} P_m^{-5/8} \, , \; {\hat k_*} \sim  S^{3/8} P_m^{-3/16} \, , \; {{\hat \delta }_{{\rm{in}}*}} \sim  S^{-5/8} P_m^{5/16} \, ,
\end{equation}
as in \citet{LSU13} and \citet{ComGra16}. 
Note that plasma viscosity leads to a decrease of the asymptotic aspect ratio ($1/{{\hat a}_*}$) and of the growth rate associated to it. As a consequence of these two factors, the validity of relations (\ref{ViscoSP_scalings}) can be extended to a larger $S$-domain, as shown in the subsequent analysis. On the other hand, for large enough Lundquist numbers, it is necessary to evaluate ${\hat a}_*$ from the inverse-aspect-ratio equation that is valid in the $P_m \gg 1$ regime, which corresponds to
\begin{equation} \label{eq_3aVISCO} 
\ln \left( {\frac{{{\lambda_{\delta} }}}{{2\hat \psi _0^{1/2}}}\frac{{\hat a_*^{1/4}}}{{{S^{1/4}}}}P_m^{1/8}} \right) = \frac{1}{2}\int_{{{\hat a}_0}}^{{{\hat a}_*}} {\hat \gamma (\hat a)\frac{{d\hat t}}{{d\hat a}}d\hat a} \, .
\end{equation}
In a manner similar to the $P_m \ll 1$ case, we first evaluate ${\hat a}_*$ and the properties of the plasmoid instability for an exponentially thinning reconnection layer, and then generalize the obtained results to also encompass algebraic thinning layers.

\subsection{Exponentially shrinking current sheet}

For strong plasma viscosity we have $\hat a_\infty = S^{-1/2}P_m^{1/4}$ \citep{Park1984} in the exponential thinning function described by Eq. (\ref{ExpCurr}). Considering the case ${\hat t} < {{\hat t}_T}$, and adopting the same approximations employed for the resistive regime, the solution of Eq. (\ref{eq_3aVISCO}) can be written as
%
\begin{equation}\label{espl_vis_a_GENERAL} 
{{\hat a}_*} \simeq \frac{{{\tau ^{2/3}}}}{{{S^{1/3}}P_m^{1/6}}}{\left( {\ln {\theta _V}} \right)^{ - 2/3}}  \, ,
\end{equation}
where 
\begin{equation}
\theta_V := \theta_R {P_m^{(1 + 2\alpha )/2}} \, .
\end{equation}
Therefore, the growth rate, wavenumber and inner layer width at the end of the linear phase become
\begin{equation} \label{espl_vis_gamma_GENERAL} 
{{\hat \gamma }_*} \simeq  \lambda_{\gamma} \frac{\ln {\theta_V}}{\tau} \, ,
\end{equation}
\begin{equation}\label{espl_vis_k_GENERAL} 
{{\hat k}_*} \simeq {\lambda_k} S^{1/6}P_m^{1/3} {\left( {\frac{{\ln {\theta_V}}}{\tau }} \right)^{5/6}} \, ,
\end{equation}
\begin{equation}\label{espl_vis_delta_GENERAL} 
{{\hat \delta }_{{\rm{in*}}}} \simeq \frac{\lambda_\delta}{S^{1/2}}  {\left( {\frac{\tau }{{\ln {\theta_V}}}} \right)^{1/2}} \, .
\end{equation}
Plasma viscosity enters in all scaling laws through the logarithmic contributions, and also as power law factors in the relations for the final aspect ratio of the current sheet and the wavenumber of the plasmoids.

For definiteness, we consider again the case in which the natural noise amplitude is approximately constant ($\alpha = 0$). In this case Eq. (\ref{espl_vis_a_GENERAL}) reduces to
\begin{equation} \label{espl_aVISC2}
{{\hat a}_*} \simeq  \frac{{{\tau ^{2/3}}}}{{{S^{1/3}}P_m^{1/6}}} {\left[ {\ln \left( {\frac{\tau P_m^{1/2}}{S^{2}}\frac{\hat a_0^{3}}{{\hat w}_0^6}} \right)} \right]^{ - 2/3}} \, .
\end{equation}
This relation indicates that an increase of the magnetic Prandtl number leads to a \emph{decrease} of the final inverse-aspect-ratio. 
\begin{figure}
\begin{center}
\includegraphics[width=8.6cm]{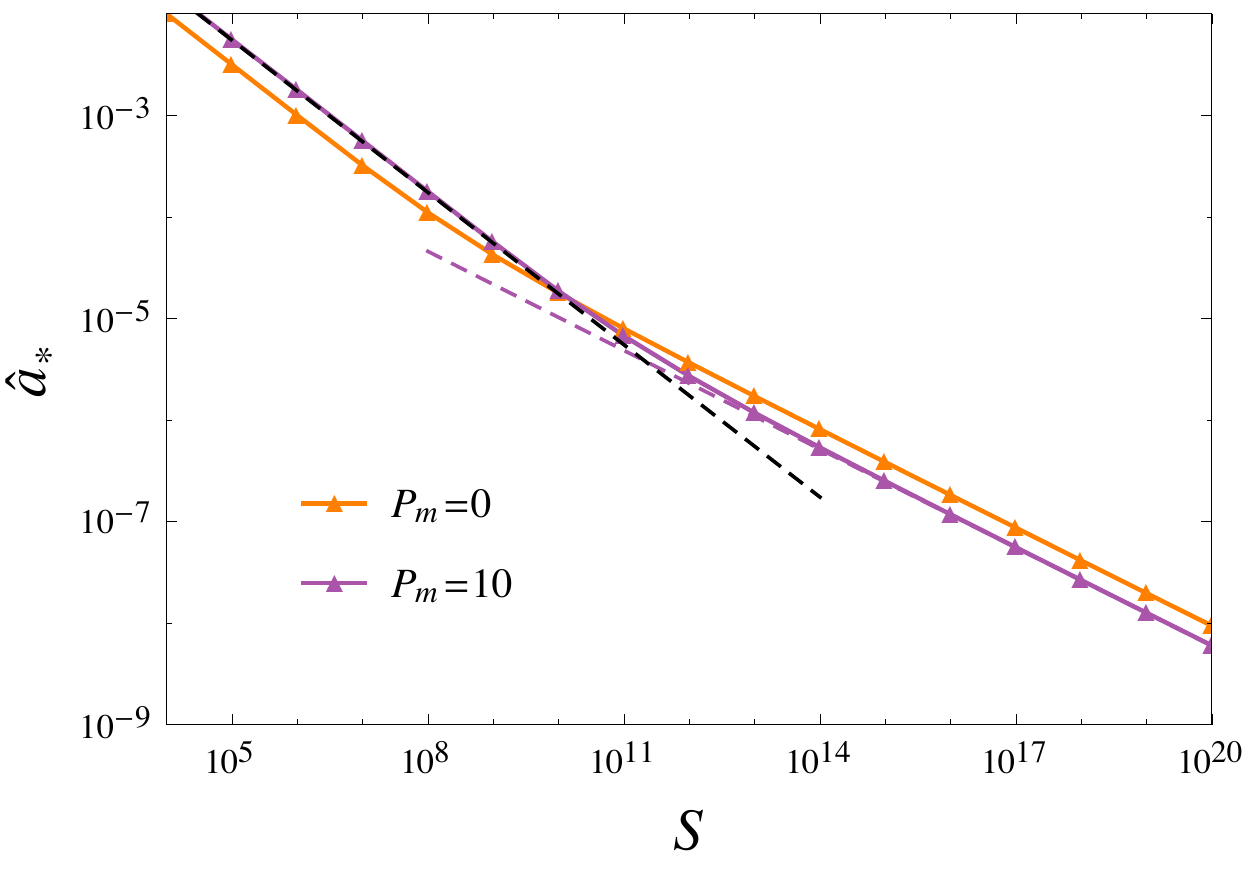}
\end{center}
\caption{Final inverse-aspect-ratio ${{\hat a}_*}$ as a function of the Lundquist number $S$ for two values of the magnetic Prandtl number $P_m$. In both cases ${\hat w}_0 = 2 ({{\varepsilon} \, {\hat a_0}})^{1/2} = 10^{-15}$, $\tau = 1$, and ${\hat a}_0 = 1/\pi$. The solid lines refer to the numerical solution of the system (\ref{Eq_1a})-(\ref{Eq_1b}). The purple dashed line refers to Eq. (\ref{espl_aVISC2}), while the black dashed line denotes the viscous Sweet-Parker scaling ${{\hat a}_*} \sim S^{-1/2} P_m^{1/4}$ with $P_m=10$.}
\label{fig5}
\end{figure}
Note that this trend is the opposite to what occurs in the viscous Sweet-Parker limit that is valid for lower $S$-values. It can be understood by noticing that plasma viscosity reduces the growth rate for a fixed value of the aspect ratio, which implies that smaller values of the inverse-aspect-ratio are required for the onset and rapid growth of the final dominant mode. The described behavior is also clearly evident from Fig. \ref{fig5}, which illustrates the final inverse-aspect-ratio as a function of $S$ for two different values of the magnetic Prandtl number.

\begin{figure}
\begin{center}
\includegraphics[width=8.6cm]{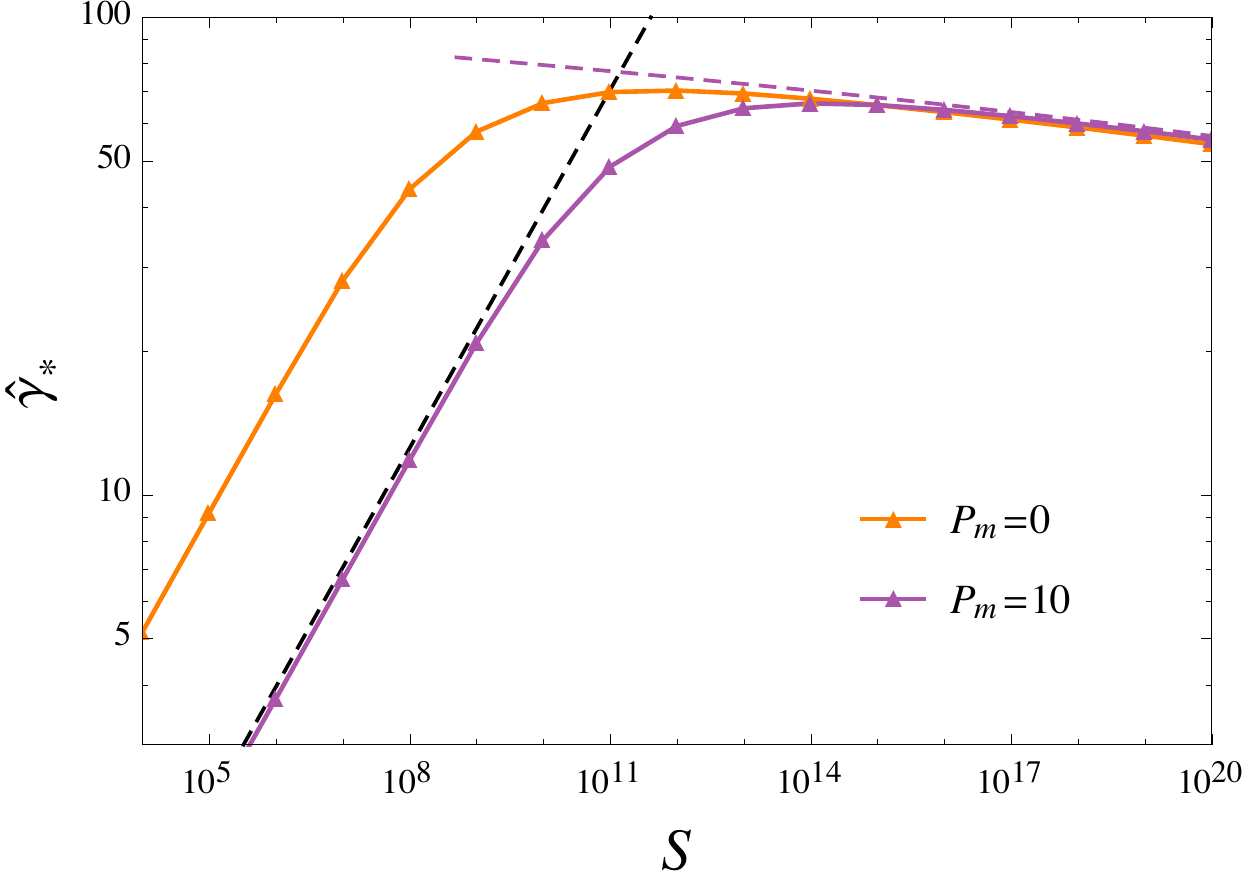}
\end{center}
\caption{Final (dominant mode) growth rate ${{\hat \gamma}_*}$ as a function of the Lundquist number $S$ for $P_m = 0$ (orange) and $P_m = 10$ (purple). Other parameters are the same as in Fig. \ref{fig5}. Solid lines refer to the numerical solution of the system (\ref{Eq_1a})-(\ref{Eq_1b}) with ${\hat \gamma}$ given by Eq. (\ref{fullgamma}). The purple dashed line refers to Eq. (\ref{EXP_gammaV}), while the black dashed line denotes the viscous Sweet-Parker based scaling ${{\hat \gamma }_*} \sim S^{1/4} P_m^{-5/8}$ with $P_m=10$.}
\label{fig7}
\end{figure}

For constant noise ${\hat \psi}_0$ we also have 
\begin{equation} \label{EXP_gammaV}
{{\hat \gamma }_*} \simeq  \frac{\lambda_{\gamma}}{\tau } \ln \left( {\frac{\tau P_m^{1/2}}{S^2}\frac{\hat a_0^3}{{\hat w}_0^6}} \right) \, ,
\end{equation}
\begin{equation} \label{EXP_kV}
{{\hat k}_*} \simeq \lambda_k \frac{S^{1/6}P_m^{1/3}}{\tau^{5/6}} {\left[ {\ln \left( {\frac{\tau P_m^{1/2}}{S^2}\frac{\hat a_0^3}{{\hat w}_0^6}} \right)} \right]^{5/6}} \, ,
\end{equation}
\begin{equation} \label{EXP_deltaV}
{{\hat \delta}_{\rm{in*}}} \simeq \lambda_{\delta} {\left( {\frac{\tau}{S}} \right)^{1/2}}{\left[ {\ln \left( {\frac{\tau P_m^{1/2}}{S^2}\frac{\hat a_0^3}{{\hat w}_0^6}} \right)} \right]^{ - 1/2}} \, .
\end{equation}
We can see that the final growth rate exhibits a weak dependence on the magnetic Prandtl number. This can be appreciated also from Fig. \ref{fig7}, where the curves for $P_m=0$ and $P_m=10$ begin to overlap at very large values of $S$. 
\begin{figure}
\begin{center}
\includegraphics[width=8.6cm]{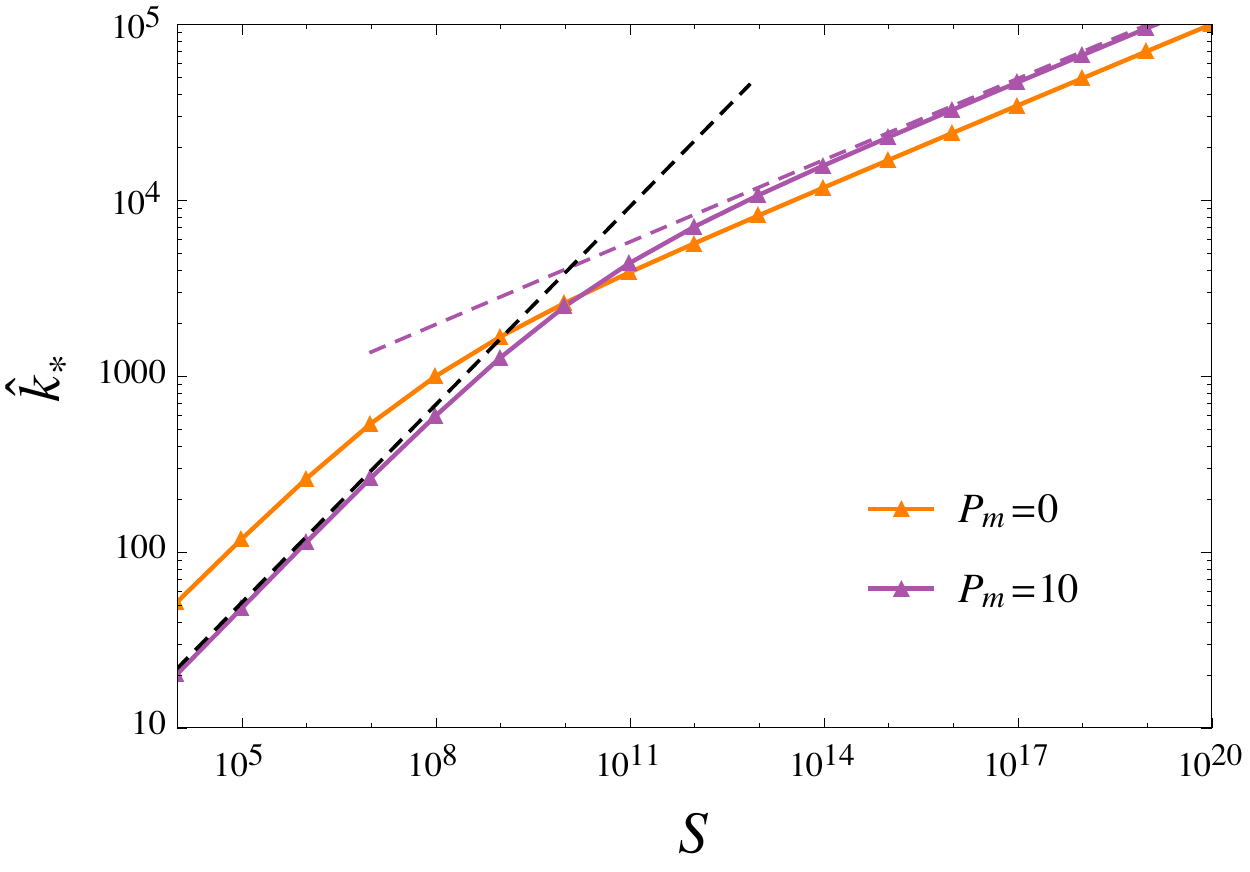}
\end{center}
\caption{Final (dominant mode) wavenumber ${{\hat k}_*}$ as a function of the Lundquist number $S$ for $P_m = 0$ (orange) and $P_m = 10$ (purple). Other parameters are the same as in Fig. \ref{fig5}. Solid lines refer to the numerical solution of the system (\ref{Eq_1a})-(\ref{Eq_1b}). The purple dashed line refers to Eq. (\ref{EXP_kV}), while the black dashed line denotes the viscous Sweet-Parker based scaling ${{\hat k}_*} \sim S^{3/8} P_m^{-3/16}$ with $P_m=10$.}
\label{fig8}
\end{figure}
\begin{figure}
\begin{center}
\includegraphics[width=8.6cm]{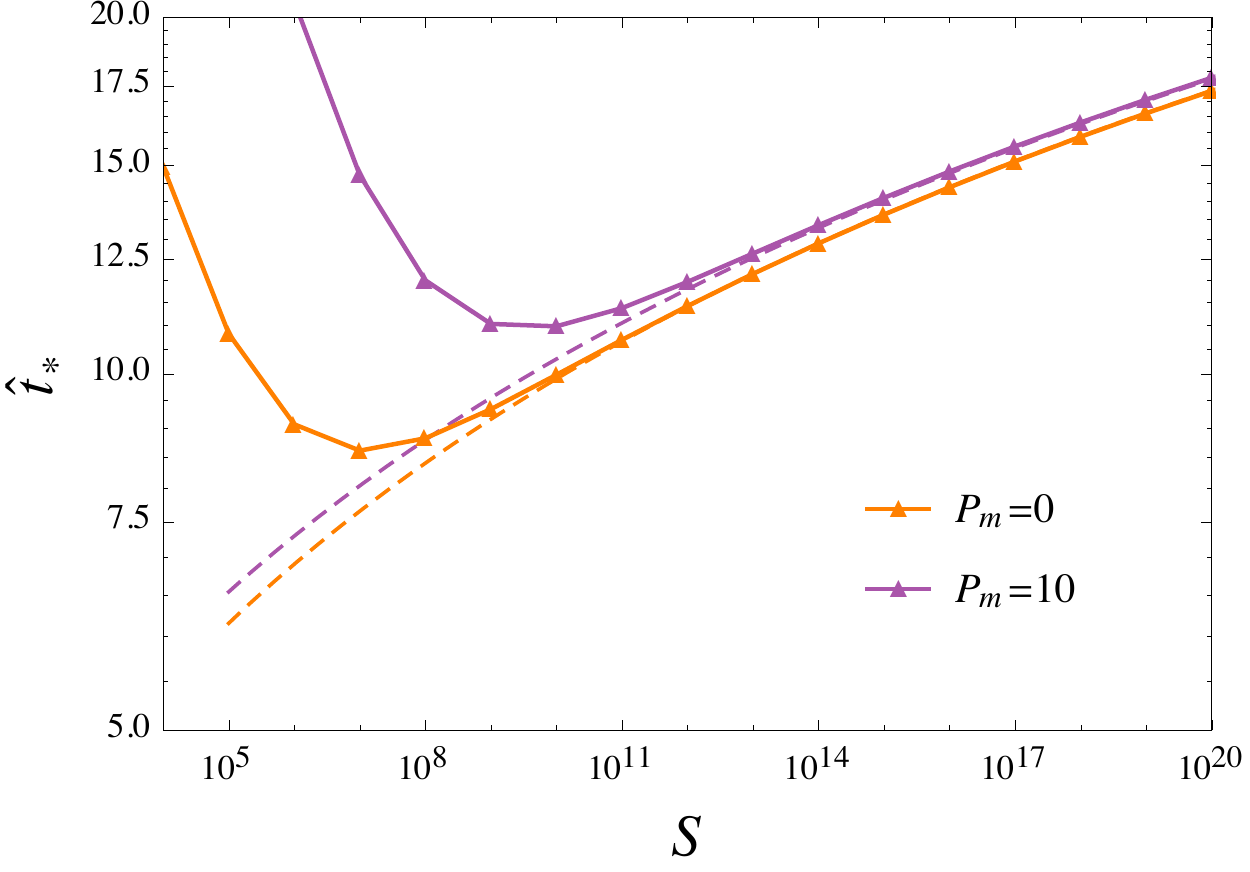}
\end{center}
\caption{Final time ${{\hat t}_*}$ as a function of the Lundquist number $S$ for $P_m = 0$ (orange) and $P_m = 10$ (purple). Other parameters are the same as in Fig. \ref{fig5}. The solid and dashed lines refer to the numerical [Eqs. (\ref{Eq_1a})-(\ref{Eq_1b})] and analytic [Eq. (\ref{fin_timeVISC})] solutions, respectively.}
\label{fig9}
\end{figure}
A weak dependence on the magnetic Prandtl number also occurs for the final inner layer width, as seen from Eq. (\ref{EXP_deltaV}). In contrast, a stronger dependence on $P_m$ is found in the final dominant mode wavenumber. In particular, Eq. (\ref{EXP_kV}) predicts an increase in the number of plasmoids for larger values of $P_m$. This behavior is verified for large values of $S$ in Fig. (\ref{fig8}). Notice that the opposite trend occurs for lower $S$ values, where ${{\hat k}_*}$ decreases as $P_m$ increases - this is evident from the second of relations (\ref{ViscoSP_scalings}) and Fig. \ref{fig8}. Thus, on interstellar scales, with high values of $P_m$ (and $S$), one would expect the production of a higher number of plasmoids compared to the resistive case.

In analogy with the resistive case, we have that the elapsed time since the beginning of the current sheet evolution is given by
\begin{equation} \label{fin_timeVISC}
{{\hat t}_*} \simeq \tau \ln \left[ {{\hat a}_0 \frac{S^{1/3}P_m^{1/6}}{\tau^{2/3}}{{\left( {\ln {\theta _V}} \right)}^{2/3}}} \right] \, ,
\end{equation}
while the intrinsic timescale of the plasmoid instability is
\begin{equation} \label{intr_timeVISC}
{\tau _p} \simeq \tau \ln \left[ {\lambda_k {{\left( {\ln {\theta _V}} \right)}^{3/2}}} \right] \, .
\end{equation}
The elapsed time ${{\hat t}_*}$ exhibits a weak dependence on the magnetic Prandtl number. In particular, Eq. (\ref{fin_timeVISC}) reveals that the total time ${{\hat t}_*}$ slightly increases for increasing $P_m$-values, which is also apparent upon inspecting Fig. \ref{fig9}. Also, the minimum of ${{\hat t}_*}$ occurs at larger $S$ since $S_T$ increases with $P_m$. The timescale $\tau_p$ displays an even weaker $P_m$-dependence to the point that it is fair to say that $\tau_p$ remains almost unchanged with respect to the results obtained for the resistive regime.

\begin{figure}
\begin{center}
\includegraphics[width=8.6cm]{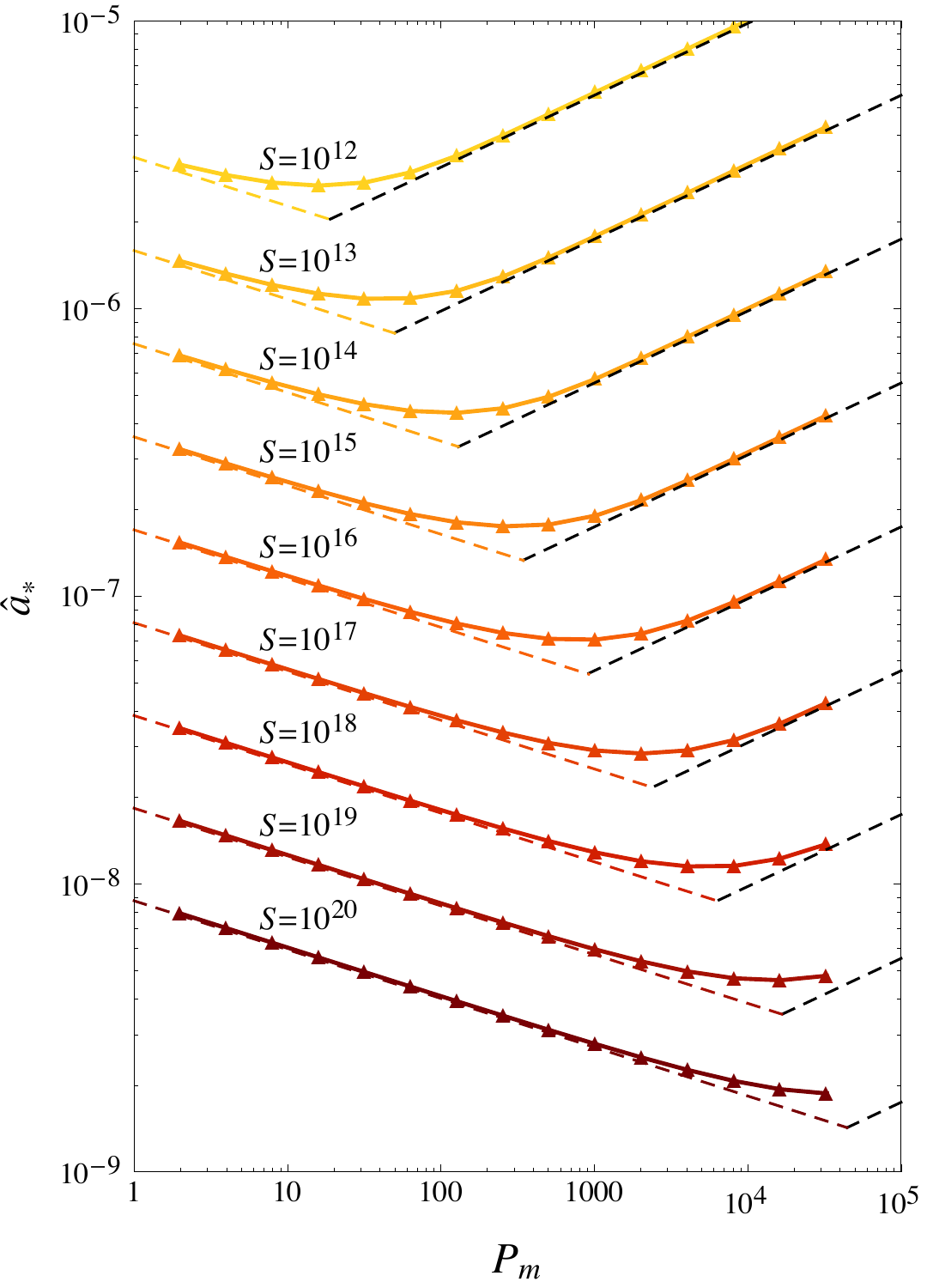}
\end{center}
\caption{Final inverse-aspect-ratio ${{\hat a}_*}$ as a function of the magnetic Prandtl number $P_m$ for different values of the Lundquist number $S$. Other parameters are the same as in Fig. \ref{fig5}. Solid lines refer to the numerical solution of the system (\ref{Eq_1a})-(\ref{Eq_1b}). The black dashed lines denote the viscous Sweet-Parker scaling ${{\hat a}_*} \sim S^{-1/2} P_m^{1/4}$, while the other dashed lines refer to Eq. (\ref{espl_aVISC2}).}
\label{fig6}
\end{figure}

An important point that must be recognized is that increasing the plasma viscosity eventually leads to the persistence of the viscous Sweet-Parker regime even for extremely large $S$-values. Hence, for a system with a given Lundquist number, there exist a value of $P_m$ for which the final inverse-aspect-ratio ${{\hat a}_*}$ is minimum. This fact is confirmed by inspecting Fig. \ref{fig6}, where the final inverse-aspect-ratio has been displayed as a function of $P_m$ for different Lundquist numbers.
The value of $P_m$ corresponding to the minimum inverse-aspect-ratio can be obtained by equating the two asymptotic behaviors of ${{\hat a}_*}$. This gives us the transitional magnetic Prandtl number
\begin{equation} \label{critical_Pm}
{P_m}_T = S^{2/5} {\left[ {\frac{{\bar \alpha }}{\tau }\,W\left( {\frac{1}{{\bar \alpha }}{{\left( {\frac{{{ (\tau /S)^{9(2 - \alpha )/10}}}}{{2^6{\varepsilon^3} }}} \right)}^{1/\bar \alpha }}} \right)} \right]^{-8/5}} \, .
\end{equation}
where $\bar \alpha := (4+8\alpha)/5$. This expression indicates that the transitional magnetic Prandtl number should increase monotonically with $S$, which is indeed confirmed from Fig. \ref{fig6}. 
Finally, we can obtain easily also the transitional Lundquist number 
\begin{equation}  \label{} 
S_T = P_m^{5/2} {\left[ {\frac{{\tilde \alpha }}{\tau }\,W\left( {\frac{1}{{\tilde \alpha }}{{\left( {\frac{{{ (\tau^2 /P_m)^{9(2-\alpha)/4}}}}{{2^6{\varepsilon^3} }}} \right)}^{1/\tilde \alpha }}} \right)} \right]^4}  \, .
\end{equation}
This relation tells us that $S_T$ has a strong dependence on $P_m$, implying that plasma viscosity can significantly extend the domain of existence of (viscous) Sweet-Parker current sheets.

\subsection{Generalized current sheet shrinking} \label{SecGenCurrShrinkVISC}
We complete the visco-resistive scalings by extending the previously obtained relations to the various current thinning possibilities described by Eq. (\ref{GenCurrShrink}) with ${{\hat a}_\infty } = S^{-1/2} P_m^{1/4}$. For ${\hat t} < {{\hat t}_T}$, using the same approximations employed for the resistive regime, the final inverse-aspect-ratio of the current sheet becomes
\begin{equation}\label{GENespl_a4VISC}
{{\hat a}_*} \sim {\left( {\frac{{\hat a_0^{2/\chi }\tau }}{{{S^{1/2}}P_m^{1/4}\ln {\Theta_V}}}} \right)^\zeta } \, ,
\end{equation}
where
%
\begin{equation}
{\Theta _V}: = {\Theta_R} {P_m^{3(4 + 2\alpha -2\zeta + 5\alpha \zeta )/16}} \, .
\end{equation}
Therefore, the final growth rate, wavenumber, and inner layer width in the visco-resistive regime are
\begin{equation}
{{\hat \gamma }_*} \sim {S^{(3\zeta  - 2)/4}}P_m^{(3\zeta  - 2)/8}{\left( {\frac{{\ln {\Theta_V}}}{{\hat a_0^{2/\chi }\tau }}} \right)^{3\zeta /2}} \, ,
\end{equation}
\begin{equation}
{{\hat k }_*} \sim {S^{(5\zeta  - 2)/8}}P_m^{(5\zeta  + 2)/16}{\left( {\frac{{\ln {\Theta_V}}}{{\hat a_0^{2/\chi }\tau }}} \right)^{5\zeta /4}} \, ,
\end{equation}
\begin{equation}
{\hat \delta}_{\rm{in*}}  \sim {S^{ - (2 + 3\zeta )/8}} P_m^{(2-3\zeta)/16}{\left( {\frac{{\hat a_0^{2/\chi} \tau}}{{\ln {\Theta_V}}}} \right)^{3\zeta/4}} \, ,
\end{equation}
and the final elapsed time is
\begin{equation}
{{\hat t}_*} \sim \frac{{\chi \tau }}{2}\left[ {\hat a_0^{(2\chi  - \zeta )/{\chi ^2}}{{\left( {\frac{{{S^{1/2}}P_m^{1/4}}}{\tau }\ln {\Theta_V}} \right)}^{2\zeta /\chi }} - 1} \right] \, .
\end{equation}
For $\chi \to \infty$ we recover the scaling relations obtained for exponentially thinning current sheets, while other choices of $\chi$ enable us to obtain the scaling laws for different cases of algebraic thinning. Note also that a particularly important and robust effect of plasma viscosity, which is common to both exponential and algebraically thinning current sheets, is the fact that it gives rise to a significant increase in the number of plasmoids that enter the nonlinear evolutionary phase for most of the astrophysically relevant (very large $S$) plasmas. 


\section{Discussion} \label{Discuss}
In this Section, we shall discuss some of the primary astrophysical consequences of the obtained scaling laws. We have shown that, in astrophysical environments, reconnecting current sheets break up before they can reach the aspect ratio predicted by the widely employed Sweet-Parker model \citep{Sweet1958,Parker1957}. The degree of discrepancy as compared to this classic prediction depends on several factors: the Lundquist number ($S$), the magnetic Prandtl number ($P_m$), the noise (both the amplitude and the spectrum) of the system ($\psi_0$), the characteristic rate of current sheet evolution ($1/\tau$), and the thinning process (taken into account by $\chi$ in our generalized current shrinking function). The importance of the noise and the thinning process have been often underestimated in previous studies. Of the two, the thinning process ($\chi$) is the one that has stronger effects in the behavior of the plasmoid instability, as can be seen from the generalized scaling laws derived in Sections (\ref{SecGenCurrShrink}) and (\ref{SecGenCurrShrinkVISC}). This points out the importance of understanding the mechanism that drives the current sheet thinning in order to apply the correct scaling relations.

We emphasize that, during the current sheet thinning, the reconnection rate remains slow. It is only after the plasmoids become nonlinear and break up the reconnecting current sheet that a sudden increase in the reconnection rate is manifested. When fast reconnection is triggered, during the highly nonlinear evolutionary phase of the plasmoids, the reconnecting current sheet is replaced by a chain of plasmoids of different sizes separated by secondary current sheets \citep{ShiTa01,BHYR,Cassak2009,HB10,ULS10,LSSU12,Taka2013,CGW15,Shiba2015}. This fragmentation could reach kinetic scales, allowing an even faster Hall/collisionless reconnection regime \citep{Daugh2006,Daugh2009,Shep2010,HBS11,JD11,HB13,ComBhatt16}. Therefore, the general analysis presented in this paper enables us to determine the onset of fast magnetic reconnection by means of the plasmoid instability, while the problem of the reconnection rate during the fast reconnection regime is not addressed in this work.

Finally, we observe that the growth of the plasmoids slows down when they enter into the early nonlinear phase, which occurs for ${\hat \delta}_{\rm{in}} < {\hat w } < {\hat a }$. But, in spite of this ``deceleration'', they can complete the early nonlinear evolution in a short timescale. This is due to the fact that the nonlinear evolution of the plasmoid instability does not exhibit a slow Rutherford evolution as it occurs for nonlinear $m \geq 2$ magnetic islands in fusion devices. Indeed, a Rutherford evolution \citep{Ruther1973} requires ${{\hat \Delta}'} {\hat w } \ll 1 $, while the plasmoids enters into the nonlinear phase with ${{\hat \Delta}'} {\hat w } \gtrsim 1 $. This can be shown by using Eqs. (\ref{k_dominant}) and (\ref{delta_dominant}) for the resistive case, or Eqs. (\ref{k_dominant_VISCO}) and (\ref{delta_dominant_VISCO}) for the visco-resistive case. In both cases, at the beginning of the nonlinear phase, we obtain
\begin{eqnarray} \label{}
{{\hat \Delta}'} {{\hat w }_*} \approx 2 \, .
\end{eqnarray}
Therefore, after the linear regime the plasmoid instability will evolve only through a fast Waelbroeck phase \citep{Waelbr1989}, as detailed in \citet{ComGra16} for the case of a time-independent current sheet. An important consequence is that the end of the linear phase practically corresponds to the disruption of the reconnecting current sheet when it is defined by the condition ${\hat w} \sim {\hat a}$ \citep{ComGra16,UL16}. One could also define the disruption of the current sheet to correspond exactly to the end of the linear phase. Indeed, at the end of the linear phase, the current density fluctuations are of the same order of the background
current density, implying that the reconnecting current sheet loses its
integrity and can be regarded as having been disrupted \citep{Huang2017}.

In the following, we apply the obtained scaling relations to reconnecting current sheets in two different astrophysical systems: the solar corona (where $P_m \ll 1$) and the warm interstellar medium (where $P_m \gg 1$).

\subsection{The Solar Corona}

The solar corona is one of the most typical environments where reconnection is expected to play a fundamental role. In fact, magnetic reconnection is considered as the leading mechanism for energy release in the form of solar flares \citep{ShiM11,Barta2011a,Barta2011b,SLM11,Li2015,ShiTa16,Jan17}. It is also responsible for coronal mass ejections \citep{Mill10,Chen11,Karpen12,Ni12,Mei2012,LMS15}. Furthermore, it is thought that reconnection may play a fundamental role in heating the solar corona \citep{Klim06,PDM,DMB}. Due to the relevance of magnetic reconnection in the solar corona, we shall explore some of the predictions of our scaling laws in this context, and compare them against the scalings based on Sweet-Parker current sheets \citep{TS97,LSC07,BHYR,BBH12,HB13,LSU13,ComGra16}.

We consider a typical case of an exponentially thinning current sheet in the solar corona. In this environment, we can adopt $B_0 = 50$ G, $n = 10^{10}$ cm$^{-3}$, $T = 10^6$ K, and $L = 2 \times 10^4$ km \citep{JD11}. We assume that the density $n$ and the temperature $T$ are the same for electrons and ions. From these parameters, we obtain $v_A \approx 10^{3}$ km/s and $S \approx 10^{13}$. Since the (perpendicular) magnetic Prandtl number is low in the solar corona, we can adopt the scaling laws obtained in Sec. \ref{SecResistive}. Exponentially shrinking current sheets form on the Alfv\'enic timescale \citep{Kuls05}, enabling us to set $\tau=1$. We also set $c_k=1$ and $c_\gamma=1/2$, which is consistent to what can be found from the full solution of the principle of least time, and choose ${\hat a_0} = 1/\pi$. The latter quantity is chosen such that the starting time corresponds to the moment in which the longest mode (with $\hat k = \pi$) becomes marginally unstable. Finally, we assume a normalized perturbation $\hat{\psi}_0 = 10^{-15}$ for all wavelengths. The natural noise of the system is clearly subject to a great degree of uncertainty, however, since the scaling laws of the plasmoid instability are weakly (logarithmically) dependent on the noise level, the final results will not be very sensitive to the actual choice of $\hat{\psi}_0$. 

Expressing our final results in dimensional units, we find
\begin{eqnarray} \label{CLMod}
    && a_* \approx 8 \times 10^3 \,\mathrm{cm}, \nonumber \\
    && \gamma_* \approx 20\,\tau_A^{-1}, \nonumber \\
    && {k}_* \approx 3 \times 10^3\, L^{-1}, \nonumber \\
    && \delta_{\rm{in*}} \approx 84 \,\mathrm{cm}, 
\end{eqnarray}
where $\tau_A = L/v_A \approx 20\,\mathrm{s}$. For our choice of parameters, we have $S_T \approx 2 \times 10^7$, which validates the use of Eqs. (\ref{espl_res_a2})-(\ref{espl_res_delta2}). On the other hand, using the Sweet-Parker based scalings would have yielded the following results.
\begin{eqnarray} \label{TSMod}
    && a_* \approx 6 \times 10^2 \,\mathrm{cm}, \nonumber \\
    && \gamma_* \approx 1 \times 10^3\,\tau_A^{-1}, \nonumber \\
    && k_* \approx 7.5 \times 10^4\, L^{-1}, \nonumber \\
    && \delta_{\rm{in*}} \approx 13\,\mathrm{cm}.
\end{eqnarray}
From (\ref{CLMod}) and (\ref{TSMod}), we see that the final inverse-aspect-ratio predicted by Eq. (\ref{espl_res_a2}) is higher than the Sweet-Parker scaling by more than an order of magnitude, implying that the usual Sweet-Parker current sheets cannot form in the solar corona. Similarly, also the inner resistive layer width at the end of the linear phase is thicker than the one obtained from the Sweet-Parker based scaling. 

A significant discrepancy between the two theoretical approaches is also evident from inspecting the final growth rate and wavenumber of the instability. The growth rate based on a Sweet-Parker current sheet overestimates the actual growth rate by two orders of magnitude for this particular example. Furthermore, the dominant wavenumber that emerges from the linear phase is more than an order of magnitude lower than the Sweet-Parker based solution. This implies that a lower number of plasmoids will appear at the beginning of the nonlinear phase.

An exponentially shrinking current sheet leads to a fast disruption of the current sheet - for this example, we find $t_* \approx 11\,\tau_A$. An exponential thinning can occur for a current sheet that is being driven by an ideal MHD instability. However, much longer time scales are involved if the current sheet formation is due to other driving processes that lead to an algebraic thinning of the current sheet. For instance, this could be the case when current sheet formation is driven by the motion of the photospheric footpoints of the magnetic field lines. 

In the case of an algebraic thinning current sheet, Eqs. (\ref{GENespl_res_a4})-(\ref{GENespl_res_delta4}) reveal that the discrepancies, when compared to the Sweet-Parker based predictions, are even larger that those obtained for an exponentially thinning current sheet. This serves to highlight the importance of adopting an appropriate time-evolving current sheet ansatz when carrying out analyses of the plasmoid instability in astrophysical environments.

\subsection{The Interstellar Medium}

Another astrophysical environment where magnetic reconnection is thought to play a crucial role is the interstellar medium. The importance of magnetic reconnection stems from the fact it has been advanced as a possible heating source for the interstellar medium \citep{Parker1992,Raymond1992,Zwei99,Tanuma2001,ES04}. Furthermore, the nature of magnetic reconnection under interstellar conditions has pivotal implications for the viability of both the galactic dynamo and theories of primordial magnetic fields \citep{Zweibel1997,ZB1997,Vishniac1999,Fatima2016}. In dealing with large-scale systems such as the interstellar medium, it is important to recognize that magnetic reconnection operates on time scales that are too large to be observed directly. Therefore, it is all the more essential to apply the correct theoretical model to gain information about how reconnection occurs in this system.

Here, we consider the case of an exponentially thinning current sheet in an ionized interstellar medium. For this system, we can assume the typical parameters $B_0 = 5 \times 10^{-6}$ G, $n = 0.1$ cm$^{-3}$, $T = 10^4$ K, and $L = 10^{16}$ km \citep{Ferr2001,JD11}. The density $n$ and the temperature $T$ are considered to be the same for electrons and ions. From these parameters we have $v_A \approx 3 \times 10^{6}$ cm/s, $S \approx 10^{20}$, and $P_m \approx 10$. Therefore, we must adopt the scaling laws obtained in Sec. \ref{SecViscoRes}. We set $\lambda_k=1$ and $\lambda_\gamma=1/2$ as in the previous application, and choose $\tau=1$ and ${\hat a_0} = 1/\pi$. Finally, we are left with a noise perturbation that needs to be specified. We assume the normalized noise perturbation to be a constant $\hat{\psi}_0 = 10^{-20}$, which, as explained before, does not affect significantly the final results since they depend only weakly on the perturbation amplitude. Note that for these parameters $S_T \approx 2 \times 10^{10}$. Therefore, the scaling relations that are valid for very large $S$-values are the correct ones to be used.

From Eqs. (\ref{espl_aVISC2}) and (\ref{EXP_gammaV})-(\ref{EXP_deltaV}), expressing the final results in dimensional units, we arrive at
\begin{eqnarray} \label{CLMod2}
    && a_* \approx 1.2 \times 10^{8} \,\mathrm{km}, \nonumber \\
    && \gamma_* \approx 22 \,\tau_A^{-1}, \nonumber \\
    && k_* \approx 1 \times 10^{5}\, L^{-1}, \nonumber \\
    && \delta_{\rm{in*}} \approx 1.5 \times 10^{5} \,\mathrm{km},
\end{eqnarray}
where $\tau_A = L/v_A \approx 10^{7} \,\mathrm{yr}$. In contrast, using the scalings based on a viscous Sweet-Parker current sheet, one would have ended up with
\begin{eqnarray} \label{TSMod2}
    && a_* \approx 1.8 \times 10^{6} \,\mathrm{km}, \nonumber \\
    && \gamma_* \approx 1.2 \times 10^{4} \,\tau_A^{-1}, \nonumber \\
    && k_* \approx 2 \times 10^{7}\, L^{-1}, \nonumber \\
    && \delta_{\rm{in*}} \approx 6.5 \times 10^{3} \,\mathrm{km}.
\end{eqnarray}
Due to the large Lundquist number of the system, the discrepancies with the viscous Sweet-Parker based predictions are very large. The final inverse-aspect-ratio is two orders of magnitude higher than the corresponding viscous Sweet-Parker value. Therefore, it is not possible to even come close to attaining such current sheets in this system. The inner visco-resistive layer width is also more than one order of magnitude thicker that the one obtained assuming a viscous Sweet-Parker current sheet. 

The plasmoid instability at the end of the nonlinear phase is characterized by an instantaneous growth rate that is very similar to the value obtained for the solar corona. It differs by three orders of magnitude from the predictions obtained by assuming a Sweet-Parker sheet, which serves to highlight the fact that the latter scalings are highly inapplicable in the interstellar medium. Lastly, a difference of more than two orders of magnitude is found for the number of plasmoids produced at the end of the nonlinear phase; once again, the Sweet-Parker based theory overestimates the actual number.

The exponential thinning of the current sheet leads to the final aspect ratio in a time $t_* \approx 17\,\tau_A$ for this application. After this time period has elapsed, fast reconnection can occur at the reconnection rate of $\sim10^{-2} {\left( {1+P_m} \right)}^{-1/2} v_{A} B_0$ \citep{CGW15,ComGra16}. Since the Alfv\'enic timescale is extremely large in this system, we conclude that the final aspect ratio is reached over a period of time that is $\sim 1/100$ the age of the Universe. On the other hand, a much longer time is expected to occur for algebraic thinning of the current sheet. In other words, there exists a very long period of time over which the energy build up occurs, conceivably even on the order of the Hubble time. 

\section{Conclusions} \label{SecConc}

As described in Sec. \ref{SecIntro}, the plasmoid instability has a great impact in astrophysical systems, ranging from solar flares \citep{ShiTa01,ShiM11} and coronal mass ejections \citep{Mei2012,LMS15} to blazar emissions \citep{SPG15,PGS16} and pulsar wind nebulae \citep{SirSpi14,Guo2015,SGP16}. The importance of the plasmoid instability arises from its capacity to prevent highly elongated current sheets from forming. Hence, in breaking up the current sheet, it facilitates fast magnetic reconnection, i.e. a rapid release of energy that is commensurate with most of the aforementioned astrophysical phenomena.

Despite its ubiquity and importance in many astrophysical environments, there are several unresolved issues pertaining to its dynamical evolution. These issues are not merely theoretical ones, since they can enable us to understand and predict the conditions under which fast reconnection will occur, for e.g. the timescales involved. Unlike most of the previous works \citep{TS97,LSC07,BHYR,BBH12,LSU13,ComGra16} which focused on stationary Sweet-Parker current sheets, we consider dynamically evolving current sheets. One of our most important results is that the Sweet-Parker based scalings are invalid for a large majority of space and astrophysical plasmas which are typically characterized by very high values of $S$. In this domain, the effects of the outflow in the reconnection layer become negligible and thus do not need to be taken into account \citep{NGH10,Huang2017}, implying that the validity of the new scaling relations is preserved.

The dynamical picture of the plasmoid instability is very complex since different modes become unstable at different times, and are then subject to rapid growth at different rates. Thus, in order to determine the scaling laws of the current sheet and the plasmoid instability at the end of the linear stage, we utilize a principle of least time that was delineated in a recent Letter \citep{CLHB16}. We calculate the growth rate, wavenumber, and inner layer width at the end of the linear phase, as well as the final aspect ratio of the reconnecting current sheet and the total time elapsed from a given time of its evolution. The analysis is carried out for both resistive and visco-resistive plasmas, since many well-known astrophysical plasmas fall within these regimes.

One of the most important results is that the scaling laws of the plasmoid instability are no longer simple power laws. These scaling relations depend on the Lundquist number ($S$), the magnetic Prandtl number ($P_m$), the noise (both the amplitude and the spectrum) of the system ($\psi_0$), the characteristic rate of current sheet evolution ($1/\tau$), and the thinning process ($\chi$).
We validated the obtained analytical scaling relations by comparing them against the full numerical solutions of the principle of least time, and demonstrated that the two are in excellent agreement. Furthermore, we have shown that the plasmoid instability comprises of a relatively long period of quiescence followed by rapid growth over a shorter timescale. This is important because it enables us to precisely estimate when the plasmoid instability becomes nonlinear and disrupts the reconnecting current sheet. 

We contrast the new scalings against those derived using a Sweet-Parker equilibrium by considering two different astrophysical systems. The first is the solar corona, where (perpendicular) $P_m \ll 1$, enabling us to use the resistive scalings. We show that the final aspect ratio of the current sheet, the number of plasmoids produced and the growth rate at the end of the linear stage are much lower (by 1-2 orders of magnitude) than those obtained by assuming a Sweet-Parker current sheet. As a matter of fact, the latter theory is not applicable to the solar corona, which has a very high value of $S \sim 10^{13}$. In addition, we compute the associated timescale, and show that the duration of the linear phase is about 10 times the Alfv\'en time for an exponentially thinning current sheet. 

We also considered the warm interstellar medium, which has (perpendicular) $P_m \gg 1$ and $S \sim 10^{20}$. The former implies that the visco-resistive scalings must be used since the magnetic Prandtl number is not negligible. As in the resistive case, we find significant differences compared to scalings based on viscous Sweet-Parker current sheets, since the warm interstellar medium does not fall under their domain of applicability.\footnote{However, it must be noted that an important effect of viscosity is to extend the validity of the (viscous) Sweet-Parker based scalings, albeit not to extremely high values of $S$.} We also find that the linear stage of the instability is $\gtrsim 1/100$ the Hubble time, which implies that fast magnetic reconnection cannot occur (in the considered system) before this time period. 

To summarize, we have rendered a dynamical picture of the linear stage of the plasmoid instability in general, time-evolving current sheets for both the resistive and visco-resistive cases. An important outcome of our analysis is that the earlier Sweet-Parker based scalings have a limited domain of validity and are thus not applicable to the majority of the astrophysical systems. We have derived new scaling relations that are no longer simple power laws and exhibit a complex dependence on the various parameters. By applying these scaling relations to the solar corona and the warm interstellar medium, we have highlighted some of the main advantages of the adopted theoretical framework. We anticipate that future studies can gainfully employ these scaling relations to obtain a detailed characterization of the plasmoid instability and the onset of fast reconnection in different astrophysical systems.


\acknowledgments
It is a pleasure to acknowledge fruitful discussions with Fatima Ebrahimi, Eero Hirvijoki, Hantao Ji, Russell Kulsrud, Roscoe White and Yao Zhou. This research was supported by the NSF Grant Nos. AGS-1338944 and AGS-1460169, and by the DOE Grant No. DE-AC02-09CH-11466.


\end{document}